\documentclass[aip,jcp,preprint]{revtex4-1}
\usepackage[T1]{fontenc}
\usepackage[latin9]{inputenc}
\usepackage{color}
\usepackage{amsmath}
\usepackage{graphicx}
\usepackage[unicode=true,pdfusetitle,
 bookmarks=true,bookmarksnumbered=false,bookmarksopen=false,
 breaklinks=true,pdfborder={0 0 0},pdfborderstyle={},backref=false,colorlinks=true]
 {hyperref}
\hypersetup{
 citecolor = blue, linkcolor = blue,  urlcolor  = blue}

\makeatletter

\providecommand{\tabularnewline}{\\}

\usepackage{refstyle}
\usepackage{url}
\usepackage{xcolor}

\makeatother

\begin{document}

\title{Probabilistic performance estimators for computational chemistry
methods: \\
the empirical cumulative distribution function of absolute errors }

\author{Pascal Pernot}

\affiliation{Laboratoire de Chimie Physique, UMR8000 CNRS/Univ. Paris-Sud, F-91405
Orsay, France}
\email{pascal.pernot@u-psud.fr}

\author{Andreas Savin}

\affiliation{Laboratoire de Chimie Théorique, CNRS and UPMC Univ. Paris 06, Sorbonne
Universités, F-75252 Paris, France}
\email{andreas.savin@lct.jussieu.fr}

\begin{abstract}
Benchmarking studies in computational chemistry use reference datasets
to assess the accuracy of a method through error statistics. The commonly
used error statistics, such as the mean signed and mean unsigned errors,
do not inform end-users on the expected amplitude of prediction errors
attached to these methods. We show that, the distributions of model
errors being neither normal nor zero-centered, these error statistics
cannot be used to infer prediction error probabilities. To overcome
this limitation, we advocate for the use of more informative statistics,
based on the empirical cumulative distribution function of unsigned
errors, namely (1) the probability for a new calculation to have an
absolute error below a chosen threshold, and (2) the maximal amplitude
of errors one can expect with a chosen high confidence level. Those
statistics are also shown to be well suited for benchmarking and ranking
studies. Moreover, the standard error on all benchmarking statistics
depends on the size of the reference dataset. Systematic publication
of these standard errors would be very helpful to assess the statistical
reliability of benchmarking conclusions.
\end{abstract}
\maketitle
\newpage{}

\section{Introduction}

There is a wide gap between the information provided by benchmarking
studies of computational chemistry (CC) methods and the information
needed by end-users to choose a method adapted to their specific,
application-dependent, requirements. It has been recently proposed
that an unequivocal criterion matching both aims would be the \emph{prediction
uncertainty} \cite{Pernot2015,Proppe2017}, which should enable to
infer intervals around the predicted value in which the true value
is expected to lie with a high probability \cite{Ruscic2014}. This
would indeed be a valuable benchmarking and ranking criterion (the
smaller, the better), and an essential information for users to select
an adequate method (other rational criteria being, for instance, method
availability and computational cost). 

Prediction uncertainty is not always easy to estimate and requires
a careful analysis of prediction errors, which are a mixture of modeling
errors (method), discretization errors (basis set, grid...), numerical
errors (floating-point arithmetic, convergence thresholds, stochastic
algorithms...), with an added contribution of parametric uncertainty
for semi-empirical methods \cite{Irikura2004,Cailliez2011}. Model
choice and discretization are mainly inducing systematic errors \cite{Dunning2000,Pernot2015},
while numerical and parametric sources are generally assumed to contribute
randomly. For deterministic CC methods, numerical and parametric uncertainties
are typically much smaller than systematic errors due to their founding
approximations and discretization schemes \cite{Irikura2004,Pernot2015,Pernot2017b}. 

Estimation of prediction uncertainty requires the calculated values
to be corrected, as well as possible, from systematic errors \cite{GUM}.
This is achieved, for instance, by composite methods \cite{Pople1989,Raghavachari2015},
\emph{a posteriori} correction estimated from trends in a calibration
errors set \cite{Pernot2015,Simm2016}, or machine learning \cite{Ramakrishnan2015,Rupp2015}.
Prediction uncertainty is therefore expected to quantify the \emph{unpredictable}
part of prediction errors, which is observed in the residual errors
after correction. Note that corrections are popular for some observables,
such as vibrational frequencies, much less for other ones, such as
atomization energies, and end-users most often use uncorrected results. 

Current CC methods do not generally provide estimations of their prediction
uncertainty, at the exception of the semi-empiric mBEEF density functional
approximation (DFA) and its relatives \cite{Wellendorff2014,Proppe2016,Aldegunde2016}.
Even in this case, uncertainty estimation is based on the absorption
of systematic errors into parametric uncertainty, the so-called \emph{parameter
uncertainty inflation} \cite{Pernot2017b}, an approach which has
recently be shown to be biased \cite{Pernot2017b}. Moreover, it is
practically impossible to derive a prediction uncertainty from the
usual statistics provided in the validation and ranking studies of
uncorrected CC methods \cite{Pernot2015}.

In the majority of validation and ranking (benchmarking) studies,
reference datasets are used to assess the accuracy of a method. The
quality of the reference datasets is central to this approach, and
several factors tend to limit the quantity of available data, notably
experimental ones. For instance, Karton et al.\cite{Karton2011} justify
their use of high-accuracy calculated data instead of experimental
ones by the following limitations: possibly large measurement uncertainties;
secondary contributions not included in approximate models; partial
and uneven coverage of the chemical universe, and small incentive
to the production of new data. 

In any case, the conclusions drawn from such benchmarking studies
are only valid in a \emph{statistical} sense. Summary statistics are
used to condense benchmark data and facilitate the decision of using,
or not, a given method. The most popular statistic is the mean absolute
error (MAE), which appears under various names \cite{Pernot2015},
for instance average absolute deviation (AAD) \cite{Curtiss1991}
or mean unsigned error (MUE) \cite{Peverati2014,Wang2017b}. The MUE\footnote{We adopt this acronym in the present study to avoid confusions with
atomization energies (AE), used in the application part.} is extensively used to assess and compare the performances of DFAs
\cite{Wang2017b}, but, as shown below, it might be unfit to enable
end-users to estimate the adequacy of a method for a given task. Note
that other statistics could be used and preferred to rank CC methods,
but most suffer from the same shortcomings as the MUE \cite{Civalleri2012,Savin2015}.

The aim of the present paper is to advocate the use of indicators
based on probabilistic considerations, which enable to implement user-defined
requirements for CC methods. As most benchmark studies deal with uncorrected
methods, one will consider only raw error sets. The basic idea is
to look for connections between a required accuracy and the probability
to obtain such an accuracy with a given method. In practice, one can
either specify the accuracy and check from the benchmark dataset if
the probability of getting acceptable results is high enough, or inversely,
specify a probability (as a confidence or success level) and decide
if the corresponding accuracy fits one's needs.

The probabilistic estimators are defined in Section~\ref{sec:Probabilistic-statistics-of}.
The dataset and the distributions of errors are exposed and explored
in Section~\ref{sec:Exploring-the-data}. In Section~\ref{sec:Results-and-discussion},
we show how the non-normality of the error distributions affects the
use of MUE to infer prediction error probabilities, and we develop
the application of the probabilistic estimators to a study dataset.
In order to illustrate our propositions, we consider the errors produced
by a set of DFAs on the atomization energies of the molecules in the
widely used G3/99 database \cite{Curtiss2000}. Note that it is not
the aim of this paper to recommend, or discourage, the use of a given
DFA, but only to exemplify how the indicators we propose might be
used. The Conclusion section provides recommendations for a generalized
use of probabilistic estimators.

\section{Probabilistic statistics of error distributions\label{sec:Probabilistic-statistics-of}}

In this section, we propose statistics that might help end-users to
assess the risks, in terms of prediction errors, involved with choosing
a given model approximation (\emph{e.g.}, DFA/basis-set). Our aim
is to answer two questions: for a molecule with similar properties
to the ones in the reference set
\begin{itemize}
\item what is the probability to achieve a chosen maximal error for a given
approximation?
\item what is the largest error one can expect with a chosen high confidence
for a given approximation?
\end{itemize}
Beforehand, we review basic information about distributions of errors,
considering that, for deterministic and uncorrected CC methods, these
are typically dominated by modeling and discretization errors. After
showing that modeling errors are not necessarily normally distributed
(Section~\ref{sec:Non-normality-of-model}), we introduce essential
notations and definitions of the statistics used in this study and
their estimators (Section~\ref{subsec:Notations-and-definitions}).
The ambiguity of the MUE as a probabilistic indicator is demonstrated
on the example of the folded normal distribution (Section~\ref{sec:Properties-of-the-Folded}).
Finally the probabilistic statistics proposed to complement the MUE
are presented (Section~\ref{subsec:Probabilistic-estimators}).

\subsection{Non-normality of model error distributions\label{sec:Non-normality-of-model}}

In order to illustrate the effect of a model approximation on error
distributions, let us characterize the system chosen by a number $x$
between 0 and 1. Let the property to be described depend on $x$ as
$y(x)=(1+x)^{2}$, and consider an approximation for it $\tilde{y}(x)=1+mx$,
where $m$ is a parameter chosen by some criterion. For example, 
\begin{itemize}
\item $m=2$ ensures that the property $(1+x)^{2}=1+2x+\dots$ is correctly
described for small $x$, 
\item $m=3$ guarantees that the property is exactly reproduced at the ends
of the interval ($x=0$ and $x=1$) 
\item $m=2.75$ is obtained by a least-squares fit, \emph{i.e.}, by choosing
$m$ to minimize\\
 $\int_{0}^{1}\left(\tilde{y}(x)-y(x)\right)^{2}\thinspace dx$. 
\end{itemize}
We will limit our discussion to $2<m<3$.

\begin{figure}[!t]
\noindent \centering{}\includegraphics[width=0.9\textwidth]{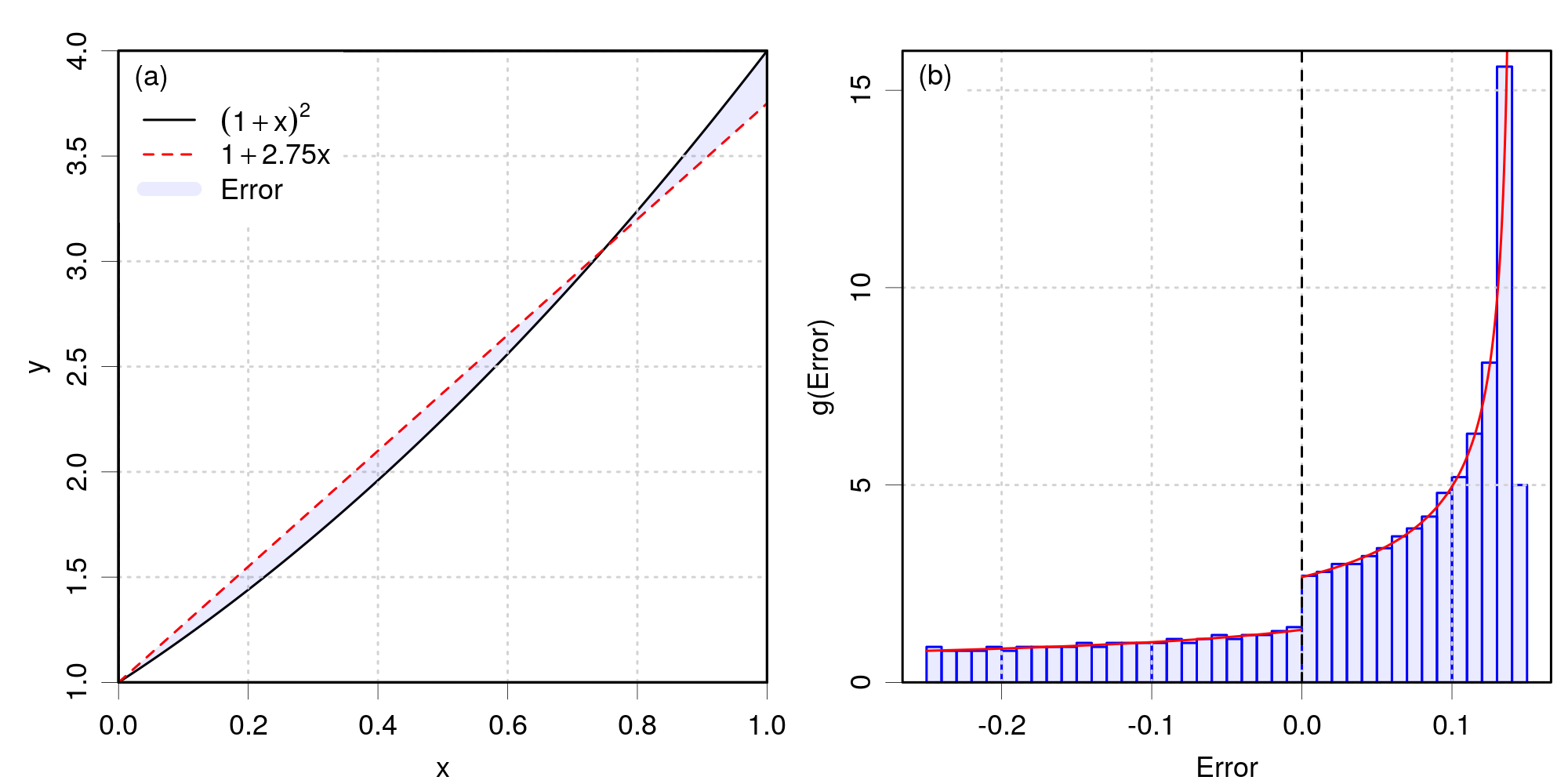}\caption{\label{fig:Model-errors}\textbf{Model error distribution for the
least-squares approximation of the curve $y=(1+x)^{2}$ by the linear
model $\tilde{y}=1+2.75x$.} (a) the curves and errors on $x\in[0,1]$;
(b) the probability density of errors $g(Error)$ (red curve) and
a histogram for a uniform sample of $x$. }
\end{figure}

Let us assume that $x$ is uniformly distributed on $\left[0,1\right]$,
\emph{i.e.}, ``the systems are chosen at random''. We would like
to know how the errors of the approximation, $e(x)=\tilde{y}(x)-y(x)$,
are distributed. If the random variable $x$ has the probability distribution
function $f(x)$ (uniform, $f(x)=1$ in our case), that of $e$, $g(e)$,
can be obtained from\cite{Taylor1997} 
\begin{equation}
g(e)=\left|\frac{dx}{de}\right|f(e)
\end{equation}
However, $e(x)$ is not monotonic on the interval of $x$ considered
here: $e(x)$ has a maximum at $x=(m-2)/2$. To obtain monotonic functions,
we subdivide the interval (0,1) into two regions, left and right of
this maximum. For each of the intervals we get 
\begin{equation}
\left|\frac{dx}{de}\right|=1/\sqrt{(m-2)^{2}-4y}
\end{equation}
However, we have to count twice the positive contributions (from the
branch $0<x<(m-2)/2$, and from $(m-2)/2<x<m-2)$, and obtain 
\begin{equation}
g(e)=\left\{ \begin{array}{ll}
1/\sqrt{(m-2)^{2}-4e} & \mbox{if \ensuremath{m-3<e<0}}\\
2/\sqrt{(m-2)^{2}-4e} & \mbox{if \ensuremath{0<e<\frac{1}{4}(m-2)^{2}}}
\end{array}\right.
\end{equation}
Evidently, this distribution of errors has nothing to do with a normal
distribution (Fig.~\ref{fig:Model-errors}). 

Even if many error sets present distributions that are less symptomatic
than the one shown here (see for instance those in Section~\ref{sec:Exploring-the-data},
Fig.~\ref{fig:fig1}), there is no reason to presume that they should
be normally distributed. They could for instance present tails with
a slow, sub-exponential, decay (so-called \emph{heavy tails}) that
prevent the reliable estimation of some common statistics. 

\subsection{Notations and definitions\label{subsec:Notations-and-definitions}}

\subsubsection{Errors / signed errors}

The calculated value $c_{i}$, for a system $i$ in a dataset of size
$N$, differs from its reference value $r_{i}$ by an error 
\begin{equation}
e_{i}=c_{i}-r_{i}
\end{equation}
The formulae for the calculation of MUE and other statistics described
below assume that the reference data and calculated values have no
uncertainty, or uncertainties much smaller than the errors themselves.
This is an ubiquitous assumption in the CC methods benchmarking literature.
In the presence of non-negligible uncertainties with heterogeneous
amplitudes, one should consider the use of weighted statistics \cite{Bevington1992}.

Considering that the errors have a probability density function (PDF),
noted $\pi(e)$, one defines the errors mean, $\mu$, standard deviation,
$\sigma$, and cumulative distribution function (CDF), $G$, by 
\begin{align}
\mu & =\int_{-\infty}^{\infty}x\,\pi(x)dx\\
\sigma & =\int_{-\infty}^{\infty}(x-\mu)^{2}\pi(x)dx\\
G(\eta) & =\int_{-\infty}^{\eta}\pi(x)\,dx
\end{align}

The CDF provides the probability that $e$ is smaller than a threshold
$\eta$: $P(e\le\eta)=G(\eta)$, where $P(X)$ is the probability
of event $X$. Inversely, the value $\eta_{p}$ below which $e$ lies
with probability $p=P(e\le\eta_{p})$, is given by the inverse of
the CDF (the quantile function), $\eta_{p}=G^{-1}(p)$.

Due to the finite size of the errors sample, one has only access to
estimates of these properties, noted with a hat (\emph{e.g.} $\hat{\mu}$)
\begin{align}
\hat{\mu} & \equiv MSE=\frac{1}{N}\sum_{i=1}^{N}e_{i}\label{eq:MSE}\\
\hat{\sigma} & \equiv RMSD=\sqrt{\frac{1}{N-1}\sum_{i=1}^{N}(e_{i}-\hat{\mu})^{2}}\label{eq:RMSD}\\
\hat{G}(\eta) & =\frac{1}{N}\sum_{i=1}^{N}\mathbf{1}_{e_{i}\le\eta}
\end{align}
where MSE is the mean signed error, RMSD is the root mean square deviation
of errors, and $\mathbf{1}_{X}$ is the indicator function of event
$X$. $\hat{G}(.)$ is called the \emph{empirical} cumulative distribution
function (ECDF). 

\subsubsection{Absolute / unsigned errors}

The absolute values of errors, or unsigned errors, $\epsilon_{i}=\left|e_{i}\right|$,
have a probability density function which results from the \emph{folding}
of $\pi(e)$ (Fig.~\ref{fig_folded}\,(a)) and is noted $\pi_{F}(\epsilon$).
The mean, standard deviation of the folded distribution and its cumulative
distribution function are 
\begin{align}
\mu_{F} & =\int_{0}^{\infty}x\,\pi_{F}(x)dx\\
\sigma_{F} & =\int_{0}^{\infty}(x-\mu_{F})^{2}\pi_{F}(x)dx\\
G_{F}(\eta) & =\int_{0}^{\eta}\pi_{F}(x)\,dx
\end{align}
and they are estimated by
\begin{align}
\hat{\mu}_{F} & \equiv MUE=\frac{1}{N}\sum_{i=1}^{N}\epsilon_{i}\\
\hat{\sigma}_{F} & =\sqrt{\frac{1}{N-1}\sum_{i=1}^{N}(\epsilon_{i}-\hat{\mu}_{F})^{2}}\\
\hat{G}_{F}(\eta) & =\frac{1}{N}\sum_{i=1}^{N}\mathbf{1_{\epsilon_{i}\le\eta}}\label{eq:ECDFU}
\end{align}

For simplicity, specific notations are used in the following for the
cumulative probabilities and \emph{percentiles} of the unsigned error
distribution
\begin{align}
C(\eta) & =\hat{G}_{F}(\eta)\label{eq:def_Pr}\\
Q_{n} & =\hat{G}_{F}^{-1}(n/100)\label{eq:def_Qn}
\end{align}
where $n$ is an integer between 0 and 100 and $n/100$ is the corresponding
probability.

\subsubsection{Statistical uncertainty of the estimators\label{subsec:Statistical-uncertainty-of}}

Due to the limited size of the benchmark datasets one has to consider
the statistical uncertainty (standard error) attached to the estimators
presented above. The formulae given below are based on the asymptotic
normality of the \emph{estimators} distributions \cite{Stuart1994}.
No strong assumption is done on the underlying \emph{error} distribution,
except for the uncertainty on the mean, where the standard deviation
has to be finite.\footnote{These estimators assume that the points in errors samples are not
correlated. However, raw error samples often display systematic trends,
as observed for some DFAs when sorting the errors by the number of
atoms in the molecules (Fig. \ref{fig_Dist_AE}). These patterns corresponding
to positive serial correlations, the standard errors are expected
to be underestimated. } The formulae apply to both signed and unsigned errors by using the
corresponding statistics, and are given here for unsigned errors:
\begin{itemize}
\item the standard error of a mean error is estimated by the usual formula
\begin{align}
u_{\hat{\mu}_{F}} & =\frac{1}{\sqrt{N}}\hat{\sigma}_{F}\label{eq:uMUE}
\end{align}
\item the standard error of a cumulative probability $C(\eta)$ is given
by \cite{Stuart1994}
\begin{equation}
u_{C(\eta)}=\sqrt{\frac{C(\eta)(1-C(\eta))}{N}}\label{eq:uPr}
\end{equation}
\item the standard error of a percentile $Q_{n}$ is estimated by Kendall's
formula \cite{Stuart1994}
\begin{equation}
u_{Q_{n}}=\frac{1}{100}\sqrt{\frac{n(100-n)}{N\,\pi_{F}^{2}(Q_{n})}}\label{eq:uQ}
\end{equation}
This formula is not well adapted for high percentiles (\emph{e.g.},
$n>80$), because the estimation of the unknown PDF $\pi_{F}(.)$
in this range is typically based on few sample points. We found it
more reliable to estimate $u_{Q_{n}}$ and confidence intervals on
$Q_{n}$ by bootstrapping \cite{Efron1979} (Appendix~\ref{sec:Bootstraping}).
\end{itemize}

\subsubsection{Remarks}

The MSE is a \emph{location} or \emph{centrality} estimator, \emph{i.e.},
it is used to estimate the position of a representative value of the
sample. As such, the MSE is helpful to detect biased error distributions
(distributions for which the MSE is not small in comparison to the
RMSD of the sample), and to modulate the interpretation of the MUE.

The MUE is particularly interesting as a robust dispersion statistics
for \emph{residuals} after model regression, \emph{i.e.}, when $|MSE|\ll MUE$,
a scenario where it is much less sensitive to outliers than the root
mean square of the residuals. However, this property is often lost
when considering error distributions: in conditions where the MSE
is not negligible before the MUE, the latter is no more a \emph{dispersion}
statistics \cite{Pernot2015}. In the limit where the bias is very
large, one gets $MUE\simeq\left|MSE\right|$, \emph{i.e.} the MUE
becomes a \emph{location} statistics. Although the interpretation
of the MUE is reputed to be ``easy'' \cite{Willmott2005,Chai2014},
it is difficult to analyze in non-ideal conditions. This crucial point
is illustrated below, in Section~\ref{sec:Properties-of-the-Folded}. 

Note that for some heavy-tailed distributions\emph{ (e.g.,} Cauchy,
slash...) such statistics as the mean and/or the variance are not
defined, but the CDF and quantiles are. 

\subsection{The Folded Normal Distribution\label{sec:Properties-of-the-Folded}}

\begin{figure}[!t]
\noindent \begin{centering}
\includegraphics[width=0.9\textwidth]{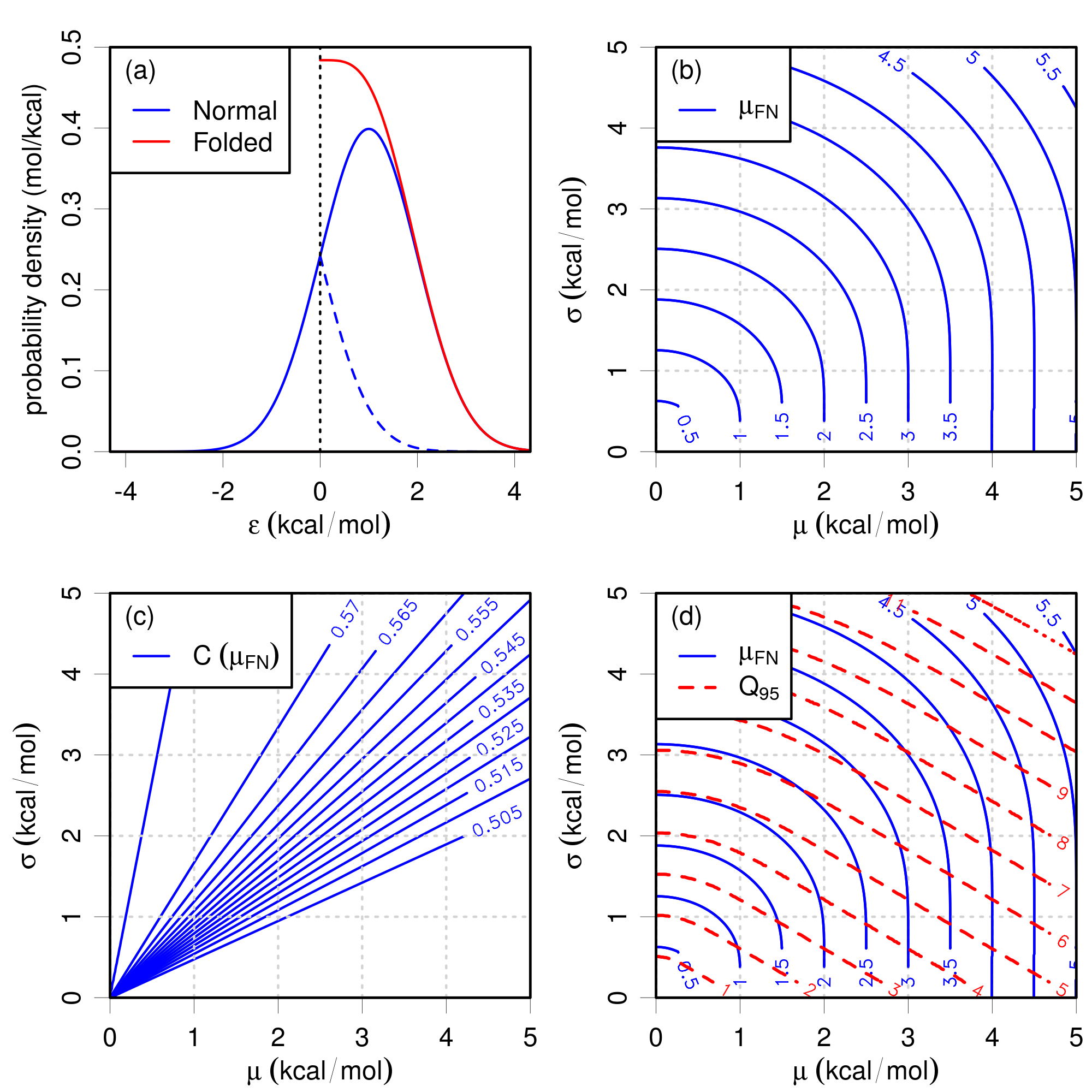}
\par\end{centering}
\noindent \centering{}\caption{\label{fig_folded}\textbf{Construction of the folded normal distribution
(FND)} \textbf{and relation of some of its properties with respect
to the mean value ($\mu$) and standard deviation ($\sigma$) of the
underlying normal distribution}: (a) Construction: the negative tail
of the normal distribution PDF (blue) is folded on the positive side
(dashed), and addition to the positive error distribution yields the
FND PDF (red); (b) Contour lines of $\mu_{FN}$, the mean of the FND,
from Eq.~\ref{eq:muF}; (c) Contour lines of the cumulative probability
$C(\mu_{FN})$, corresponding to the values of $\mu_{FN}$ in panel
(b); (d) Contour lines of the $95^{th}$ percentile of the FND, $Q_{95}$,
superimposed on the contours of $\mu_{FN}$ reported from panel (b).}
\end{figure}

If $X$ is a normally distributed random variable with mean $\mu$
and standard deviation $\sigma$, $|X|$ has a \emph{folded} normal
distribution (FND) with PDF \cite{Leone1961} (Fig.~\ref{fig_folded}\,(a))
\begin{equation}
\pi_{FN}(\epsilon;\mu,\sigma)=\frac{1}{\sqrt{2\pi\sigma^{2}}}\left[\exp\left(-\frac{\left(\epsilon-\mu\right)^{2}}{2\sigma^{2}}\right)+\exp\left(-\frac{\left(\epsilon+\mu\right)^{2}}{2\sigma^{2}}\right)\right]\label{eq:pdfF}
\end{equation}

\paragraph{Mean value.}

The mean $\mu_{FN}$ of the FND depends in a complex way on the parameters
of the original normal distribution
\begin{equation}
\mu_{FN}(\mu,\sigma)=\sigma\sqrt{\frac{2}{\pi}}\exp\left(-\frac{\mu^{2}}{2\sigma^{2}}\right)-\mu\thinspace\mathrm{erf}\left(-\frac{\mu}{\sqrt{2}\sigma}\right),\label{eq:muF}
\end{equation}
so that a same value of $\mu_{FN}$ might result from very different
normal distributions (\emph{e.g.}, small $\mu$ and large $\sigma$,
large $\mu$ and small $\sigma$). The dependence of\emph{ $\mu_{FN}$
}on\emph{ }$(\mu,\sigma)$\emph{ }is displayed by contour lines in
Fig.~\ref{fig_folded}\,(b). Note that $\lim_{\sigma\rightarrow0}(\mu_{FN})=\mu$. 

Note also that a decrease of $\mu_{FN}$ can be achieved through a
variety of paths in the $(\mu,\sigma)$ space, notably by decreasing
$\mu$ and increasing $\sigma$, or \emph{vice versa}. Therefore,
in benchmarking studies, a lower MUE does not guarantee overall better
performances, as shown in the following.

\paragraph{Cumulative probabilities.\label{subsec:Cumulative-probabilities}}

The CDF, as the integral of $\pi_{FN}$, depends also on $\mu$ and
$\sigma$ 
\begin{equation}
G_{FN}(\epsilon;\mu,\sigma)=\frac{1}{2}\left[\mathrm{erf}\left(\frac{\epsilon-\mu}{\sqrt{2}\sigma}\right)+\mathrm{erf}\left(\frac{\epsilon+\mu}{\sqrt{2}\sigma}\right)\right]\label{eq:cdfF}
\end{equation}
In order to investigate the interest of the MUE (exactly known here
as $\mu_{FN}$) as a probabilistic estimator, one can calculate the
corresponding cumulative probability 
\begin{equation}
C(\mu_{FN})=P(\epsilon\le\mu_{FN})=G_{FN}(\mu_{FN})
\end{equation}
The value depends on $\mu$ and $\sigma$ (Fig.~\ref{fig_folded}\,(c)),
and varies in the range $\left[0.5,0.5753\right]$. Even in ideal
conditions of normal error distributions, there is not a unique cumulative
probability attached to the MUE.

\paragraph{Percentiles.\label{subsec:Quantiles}}

Similarly, a chosen value of $\mu_{FN}$ corresponds to a wide range
of values for the percentiles of the folded distribution (\emph{e.g.,}
$Q_{95}$). In Fig.~\ref{fig_folded}\,(d), one can see that a single
$\mu_{FN}$ contour line crosses several contour lines for $Q_{95}$.
For instance, the $\mu_{FN}=2$ contour intersects with $Q_{95}$
lines varying in the $[2,5]$\,kcal/mol range. This shows that in
benchmarking studies, a small value of the MUE does not guarantee
good predictive performance of a method.

However, a pair of values ($\mu_{FN},\mu$) might enable to determine
a percentile uniquely. Using Fig.~\ref{fig_folded}\,(d), one can
check, for instance, that the contour line for $\mu_{FN}=2.5$, intersects
the vertical line for $\mu=2$ at a point where the value of $Q_{95}$
is about 6. This suggests that, at least for normal error distributions,
the (MUE, MSE) pair provided by many benchmark studies might be used
to infer probabilistic information on unsigned errors, in the same
way as the (MSE, RMSD) pair would on signed errors. This will be tested
in Section~\ref{subsec:Estimation-of-quantiles}.

\subsection{Probabilistic estimators\label{subsec:Probabilistic-estimators}}

We have shown above that model error distributions are not a priori
normal, and that, even for normal error distributions, the MUE cannot
provide unique probabilistic estimations. One is therefore in need
of other kind of estimators to answer the questions posed in the introduction
of this section. One needs in fact to be able to estimate probabilities
associated with a chosen error level, and/or error levels associated
with a chosen probability. The central tool for this kind of inquiry
is the CDF. As we are interested mostly in the amplitude of errors,
we will use the ECDF of \emph{unsigned} errors $\hat{G}_{F}$ (Eq.~\ref{eq:ECDFU}). 

\begin{figure}[!t]
\noindent \centering{}\includegraphics[width=0.65\textwidth]{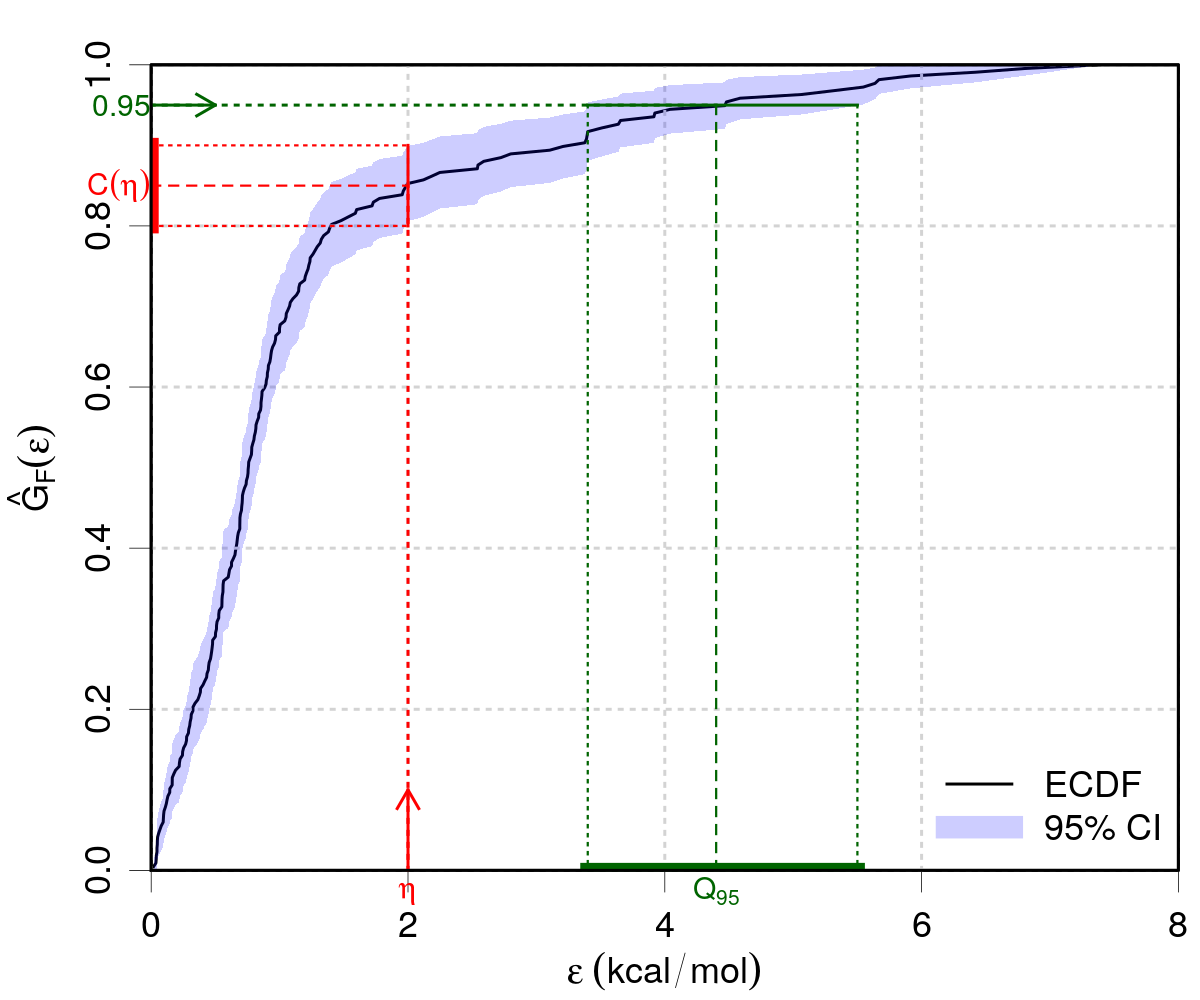}\caption{\label{fig_ECDF}\textbf{Empirical cumulative distribution function
for B3LYP absolute errors on IAE for the G3/99 set}. The shaded area
delimits the 95\% confidence interval on the ECDF. The colored lines
provide examples of the inquiries that can be done from the ECDF (see
text).}
\end{figure}

In order to be more realistic than with the FND, we illustrate the
following points on a concrete example: Fig.~\ref{fig_ECDF} shows
the ECDF of the absolute errors on intensive atomization energies
(IAE) by the B3LYP DFA. The definition of IAE is not relevant at this
stage, and is presented in Section~\ref{sec:Exploring-the-data}.
The shaded area delimits the 95\% uncertainty band on the ECDF due
to the sample size of the G3/99 dataset.

\subsubsection{Probability of obtaining acceptable results}

For the approach presented in this section, users have to decide what
is an acceptable absolute error $\eta$ for their applications. Based
on the data in the reference set, users can conclude whether their
aim of getting acceptable results can be reached. 

A trivial strategy would be to retain only methods for which $\max(\epsilon)<\eta$.
Unfortunately, as most methods present large errors for some systems,
this would hopelessly deplete the pool of usable methods. One has
thus to accept some risk, and use a probabilistic criterion. The probability
to obtain an acceptable absolute error level $\eta$ with a given
method is estimated from the ECDF as $C(\eta)$ (Eq.~\ref{eq:def_Pr}). 

As an illustration, consider Fig.~\ref{fig_ECDF}: if one chooses
an acceptance threshold for errors on IAE of $\eta=2$\,kcal/mol
(red arrow), one gets $C(2)\simeq0.85$. Considering the statistical
uncertainty on the ECDF one has indeed between 80 and 90 percent chances
to achieve this maximum error level with B3LYP. So, out of 10 calculations
for new systems with this DFA, one should expect that, on average,
only 1 or 2 will provide IAE results with errors exceeding the chosen
limit of 2\,kcal/mol.

\subsubsection{High confidence error level\label{subsec:High-confidence-error}}

Instead of obtaining the probability $C(\eta)$ after specifying the
reliability parameter $\eta$, one may decide on a required confidence
level about the outcome of a calculation and check the corresponding
error level. One specifies thus first a probability of success (\emph{e.g.},
$p=$0.90 or 0.95), and then search the largest absolute error $\epsilon$
one has to accept, so that this probability level can be reached.
Here again, the answer is given by the ECDF, through its inverse function
and the percentiles $Q_{n}$, with $n=100\times p$ (Eq.~\ref{eq:def_Qn}).

For instance, using the ECDF for B3LYP (Fig.~\ref{fig_ECDF}), the
IAE absolute error corresponding to a 0.95 probability level is $Q_{95}\simeq4.4$\,kcal/mol.
Considering the uncertainty on the ECDF, the error level to accept
lies between 3.4 and 5.5~kcal/mol. 

The risk level associated with the choice of a high-probability percentile
can also be stated in terms of the percentage of new calculations
for which the absolute errors are expected to exceed the chosen percentile.
For $Q_{n},$ this number is on average $(100-n)$\,\%. For a new
molecule with similar properties to the ones in the reference dataset,
one has on average only 5\,\% chance to exceed the $Q_{95}$ error
level. Of course, there is a distribution of excess chances, which
depends on the size of the reference dataset and on the probability
level.

For the choice of a success/risk level, one has to appreciate that,
due to the errors sample size, the uncertainty on the percentiles
increases with the probability. For B3LYP for instance, the upper
bound of a 95\% confidence interval (CI) on $Q_{95}$, noted $\left\lceil Q_{95}\right\rceil $,
is about 5.5\,kcal/mol (Table~\ref{tab:All-Stats}). If one is ready
to accept a 10\,\% risk, the values are somewhat smaller, with $Q_{90}=3.2$
and $\left\lceil Q_{90}\right\rceil =3.9$~kcal/mol. The choice of
a success/risk level has therefore to be guided by several considerations:
\begin{itemize}
\item For small reference datasets, the uncertainty on high percentiles
might be large (Appendix~\ref{sec:Bootstraping}), and it might be
pointless to discern $Q_{90}$ from $Q_{95}$. This would be the case
for datasets with less than 100 points. In the present study case,
with more than 200 points, their 95\,\% confidence intervals still
overlap (see also Table~\ref{tab:All-Stats}), but the median value
of one percentile lies outside of the 95\,\% CI of the other.
\item It is also noteworthy that higher quantiles might be more influenced
by outliers. However, a level of 10\,\% or even 5\,\% of outliers
in a dataset starts to be problematic anyway, and they should be treated
before performing statistical estimation.
\item Some heavily corrected methods, such as the composite methods for
thermochemistry, lead to quasi-normal error distributions \cite{Klippenstein2017a}.
In such cases, it has been recommended by Ruscic \cite{Ruscic2014}
to use an enlarged uncertainty $u_{95\%}$ to summarize the errors.
Using an enlarged uncertainty assumes the symmetry of the error distribution,
not its normality, and provides probabilistic information on the performance
of the method \cite{GUM}: $P(\hat{\mu}-u_{95\%}\le e\le\hat{\mu}+u_{95\%})=0.95$.
In the case of unbiased methods ($\hat{\mu}\simeq0$) this translates
for unsigned errors as $P(\epsilon\le u_{95\%})=0.95$, which is the
definition of $Q_{95}$ (Eq.~\ref{eq:def_Qn}). Therefore, by using
$Q_{95}$ as a probabilistic estimator in the case of general error
distributions, one ensures a direct link to the recommended usage
for symmetric distributions.
\end{itemize}

\section{Application}

\subsection{Exploring the data sets\label{sec:Exploring-the-data}}

To illustrate the concepts developed in this article, we consider
the errors on the atomization energies (AE) of the G3/99 database
\cite{Curtiss2000}. We base our study on published data \cite{OterodelaRoza2013},
produced with the following DFAs: PW86PBE \cite{Perdew1986,Perdew1996a},
B3LYP \cite{Lee1988,Becke1993}, PBE0 \cite{Adamo1999}, CAM-B3LYP
\cite{Yanai2004}, LC-$\omega$PBE \cite{Vydrov2006,Vydrov2006a},
PBE \cite{Perdew1996a}, BLYP \cite{Becke1988,Lee1988}, BH\&HLYP
\cite{Becke1993}, and B97-1 \cite{Hamprecht1998}. BLYP, PBE and
PW86PBE are pure functionals, the remaining are hybrids, CAM-B3LYP
and LC-$\omega$PBE using range-separation.

Due to the extensivity of the atomization energies, it has been shown
that errors typically increase with the size of the system \cite{Savin2015,Perdew2016,Margraf2017}.
To eliminate this trend, we also consider the atomization energies
per atom, noted IAE for intensive atomization energies \cite{Perdew2016}. 

\subsubsection{Benchmarking statistics}

First, we report reference statistics as found in most CC methods
benchmarking studies (Table~\ref{tab:Stats-AE}), namely, the MUE,
MSE, RMSD, Lowest Negative Error (LNE) and Highest Positive Error
(HPE). We omit the root mean squared error (RMSE, mean of the uncentered
errors) which is often reported alongside the MUE, but has no practical
interest here, and include instead the RMSD, which is useful to assess
the importance of the bias. See Section~\ref{subsec:Notations-and-definitions}
for definitions of these statistics.

\noindent 
\begin{table}[t]
\caption{\label{tab:Stats-AE}\textbf{Statistics of AE and IAE errors on the
G3/99 dataset for a selection of DFAs}. MUE: mean unsigned error;
MSE: mean signed error; RMSD: root mean square deviation; LNE: lowest
negative error; HPE: highest positive error.}

\noindent \centering{}\medskip{}
{\footnotesize{}}%
\begin{tabular}{lr@{\extracolsep{0pt}.}lr@{\extracolsep{0pt}.}lr@{\extracolsep{0pt}.}lr@{\extracolsep{0pt}.}lr@{\extracolsep{0pt}.}lr@{\extracolsep{0pt}.}lr@{\extracolsep{0pt}.}lr@{\extracolsep{0pt}.}lr@{\extracolsep{0pt}.}lr@{\extracolsep{0pt}.}lr@{\extracolsep{0pt}.}l}
\hline 
{\footnotesize{}DFA} & \multicolumn{10}{l}{{\footnotesize{}Error Statistics for AE (kcal/mol)}} & \multicolumn{2}{c}{} & \multicolumn{10}{l}{{\footnotesize{}Error Statistics for IAE (kcal/mol)}}\tabularnewline
\cline{2-11} \cline{14-23} 
 & \multicolumn{2}{c}{{\footnotesize{}MUE }} & \multicolumn{2}{c}{{\footnotesize{}MSE }} & \multicolumn{2}{c}{{\footnotesize{}RMSD}} & \multicolumn{2}{c}{{\footnotesize{}LNE }} & \multicolumn{2}{c}{{\footnotesize{}HPE}} & \multicolumn{2}{c}{} & \multicolumn{2}{c}{{\footnotesize{}MUE }} & \multicolumn{2}{c}{{\footnotesize{}MSE }} & \multicolumn{2}{c}{{\footnotesize{}RMSD}} & \multicolumn{2}{c}{{\footnotesize{}LNE }} & \multicolumn{2}{c}{{\footnotesize{}HPE }}\tabularnewline
\hline 
{\footnotesize{}B3LYP } & {\footnotesize{}7}&{\footnotesize{}8 } & {\footnotesize{}7}&{\footnotesize{}2 } & {\footnotesize{}7}&{\footnotesize{}9 } & {\footnotesize{}-7}&{\footnotesize{}8 } & {\footnotesize{}39}&{\footnotesize{}5 } & \multicolumn{2}{c}{} & {\footnotesize{}1}&{\footnotesize{}2 } & {\footnotesize{}1}&{\footnotesize{}0 } & {\footnotesize{}1}&{\footnotesize{}5 } & {\footnotesize{}-3}&{\footnotesize{}9 } & {\footnotesize{}7}&{\footnotesize{}4 }\tabularnewline
{\footnotesize{}B97-1 } & {\footnotesize{}6}&{\footnotesize{}1 } & {\footnotesize{}4}&{\footnotesize{}8 } & {\footnotesize{}6}&{\footnotesize{}8 } & {\footnotesize{}-9}&{\footnotesize{}3 } & {\footnotesize{}24}&{\footnotesize{}7 } & \multicolumn{2}{c}{} & \textbf{\footnotesize{}0}&\textbf{\footnotesize{}9}{\footnotesize{} } & {\footnotesize{}0}&{\footnotesize{}5 } & {\footnotesize{}1}&{\footnotesize{}1 } & {\footnotesize{}-3}&{\footnotesize{}2 } & {\footnotesize{}4}&{\footnotesize{}6 }\tabularnewline
{\footnotesize{}BH\&HLYP } & {\footnotesize{}32}&{\footnotesize{}3 } & {\footnotesize{}32}&{\footnotesize{}2 } & {\footnotesize{}18}&{\footnotesize{}5 } & {\footnotesize{}-7}&{\footnotesize{}4 } & {\footnotesize{}83}&{\footnotesize{}4 } & \multicolumn{2}{c}{} & {\footnotesize{}4}&{\footnotesize{}8 } & {\footnotesize{}4}&{\footnotesize{}8 } & {\footnotesize{}3}&{\footnotesize{}5 } & {\footnotesize{}-3}&{\footnotesize{}7 } & {\footnotesize{}20}&{\footnotesize{}5 }\tabularnewline
{\footnotesize{}BLYP } & {\footnotesize{}11}&{\footnotesize{}4 } & {\footnotesize{}7}&{\footnotesize{}5 } & {\footnotesize{}12}&{\footnotesize{}9 } & {\footnotesize{}-25}&{\footnotesize{}4 } & {\footnotesize{}45}&{\footnotesize{}3 } & \multicolumn{2}{c}{} & {\footnotesize{}1}&{\footnotesize{}6 } & {\footnotesize{}0}&{\footnotesize{}4 } & {\footnotesize{}2}&{\footnotesize{}2 } & {\footnotesize{}-8}&{\footnotesize{}5 } & {\footnotesize{}7}&{\footnotesize{}0 }\tabularnewline
{\footnotesize{}CAM-B3LYP } & \textbf{\footnotesize{}4}&\textbf{\footnotesize{}2 } & {\footnotesize{}2}&{\footnotesize{}3 } & {\footnotesize{}6}&{\footnotesize{}6 } & {\footnotesize{}-7}&{\footnotesize{}8 } & {\footnotesize{}32}&{\footnotesize{}7 } & \multicolumn{2}{c}{} & \textbf{\footnotesize{}0}&\textbf{\footnotesize{}9}{\footnotesize{} } & {\footnotesize{}0}&{\footnotesize{}6 } & {\footnotesize{}1}&{\footnotesize{}5 } & {\footnotesize{}-3}&{\footnotesize{}9 } & {\footnotesize{}6}&{\footnotesize{}8 }\tabularnewline
{\footnotesize{}LC-$\omega$PBE } & {\footnotesize{}5}&{\footnotesize{}1 } & {\footnotesize{}2}&{\footnotesize{}9 } & {\footnotesize{}6}&{\footnotesize{}4 } & {\footnotesize{}-14}&{\footnotesize{}0 } & {\footnotesize{}27}&{\footnotesize{}3 } & \multicolumn{2}{c}{} & {\footnotesize{}1}&{\footnotesize{}1 } & {\footnotesize{}0}&{\footnotesize{}7 } & {\footnotesize{}1}&{\footnotesize{}7 } & {\footnotesize{}-3}&{\footnotesize{}6 } & {\footnotesize{}9}&{\footnotesize{}5 }\tabularnewline
{\footnotesize{}PBE } & {\footnotesize{}18}&{\footnotesize{}9 } & {\footnotesize{}-17}&{\footnotesize{}9 } & {\footnotesize{}15}&{\footnotesize{}5 } & {\footnotesize{}-75}&{\footnotesize{}0 } & {\footnotesize{}14}&{\footnotesize{}0 } & \multicolumn{2}{c}{} & {\footnotesize{}2}&{\footnotesize{}8 } & {\footnotesize{}-2}&{\footnotesize{}5 } & {\footnotesize{}2}&{\footnotesize{}7 } & {\footnotesize{}-13}&{\footnotesize{}6 } & {\footnotesize{}2}&{\footnotesize{}8 }\tabularnewline
{\footnotesize{}PBE0 } & {\footnotesize{}5}&{\footnotesize{}5 } & {\footnotesize{}-1}&{\footnotesize{}0 } & {\footnotesize{}8}&{\footnotesize{}2 } & {\footnotesize{}-31}&{\footnotesize{}1 } & {\footnotesize{}29}&{\footnotesize{}3 } & \multicolumn{2}{c}{} & \textbf{\footnotesize{}0}&\textbf{\footnotesize{}9}{\footnotesize{} } & {\footnotesize{}0}&{\footnotesize{}2 } & {\footnotesize{}1}&{\footnotesize{}4 } & {\footnotesize{}-2}&{\footnotesize{}9 } & {\footnotesize{}6}&{\footnotesize{}5 }\tabularnewline
{\footnotesize{}PW86PBE } & {\footnotesize{}9}&{\footnotesize{}4 } & {\footnotesize{}-1}&{\footnotesize{}5 } & {\footnotesize{}12}&{\footnotesize{}2 } & {\footnotesize{}-33}&{\footnotesize{}8 } & {\footnotesize{}29}&{\footnotesize{}8 } & \multicolumn{2}{c}{} & {\footnotesize{}1}&{\footnotesize{}6 } & {\footnotesize{}-0}&{\footnotesize{}5 } & {\footnotesize{}2}&{\footnotesize{}5 } & {\footnotesize{}-11}&{\footnotesize{}3 } & {\footnotesize{}5}&{\footnotesize{}9 }\tabularnewline
\hline 
\end{tabular}{\footnotesize \par}
\end{table}

Considering AE, the DFA with the smallest MUE is CAM-B3LYP (4.2 kcal/mol).
It has a noticeable bias (MSE) of about 2.3\,kcal/mol, to be compared
to a RMSD of 6.6\,kcal/mol. The errors are dispersed in a range |HPE-LNE|
of about 40 kcal/mol. The DFA with the smallest error range in the
set is B97-1 (34\,kcal/mol), but it is more strongly biased than
CAM-B3LYP (4.8 kcal/mol) and has a larger MUE (6.1 kcal/mol). 

For intensive atomization energies, three DFAs share the lowest MUE
of 0.9 kcal/mol (B97-1, CAM-B3LYP and PBE0). Among those, PBE0 is
the least biased, but B97-1 has the smallest error range. However,
one should keep in mind that the error range might reflect the presence
of outliers, and not characterize properly the properties of the error
distribution. 

So, which DFA is the best, in the sense that it minimizes the risk
to get a large error when predicting the AE or IAE of a new system?
It is difficult to conclude from these statistics, and additional
information is clearly needed: one has to go beyond elementary summary
statistics and consider the underlying error distributions. 

\subsubsection{Error distributions in the G3/99 AE and IAE sets}

Assuming that the level of uncertainty in the reference data is negligible
(less than 1\,kcal/mol on formation enthalpies according to Curtiss
\emph{et al.} \cite{Curtiss2000}), and that the numerical errors
in the calculated data are assumed to be well controlled \cite{Irikura2004},
discrepancy between calculated and reference values in the present
dataset reflects either systematic errors from the DFA (modeling and
discretization errors) or improper reference data \cite{Pernot2015}. 

\begin{figure}[!t]
\noindent \centering{}\includegraphics[width=0.9\textwidth]{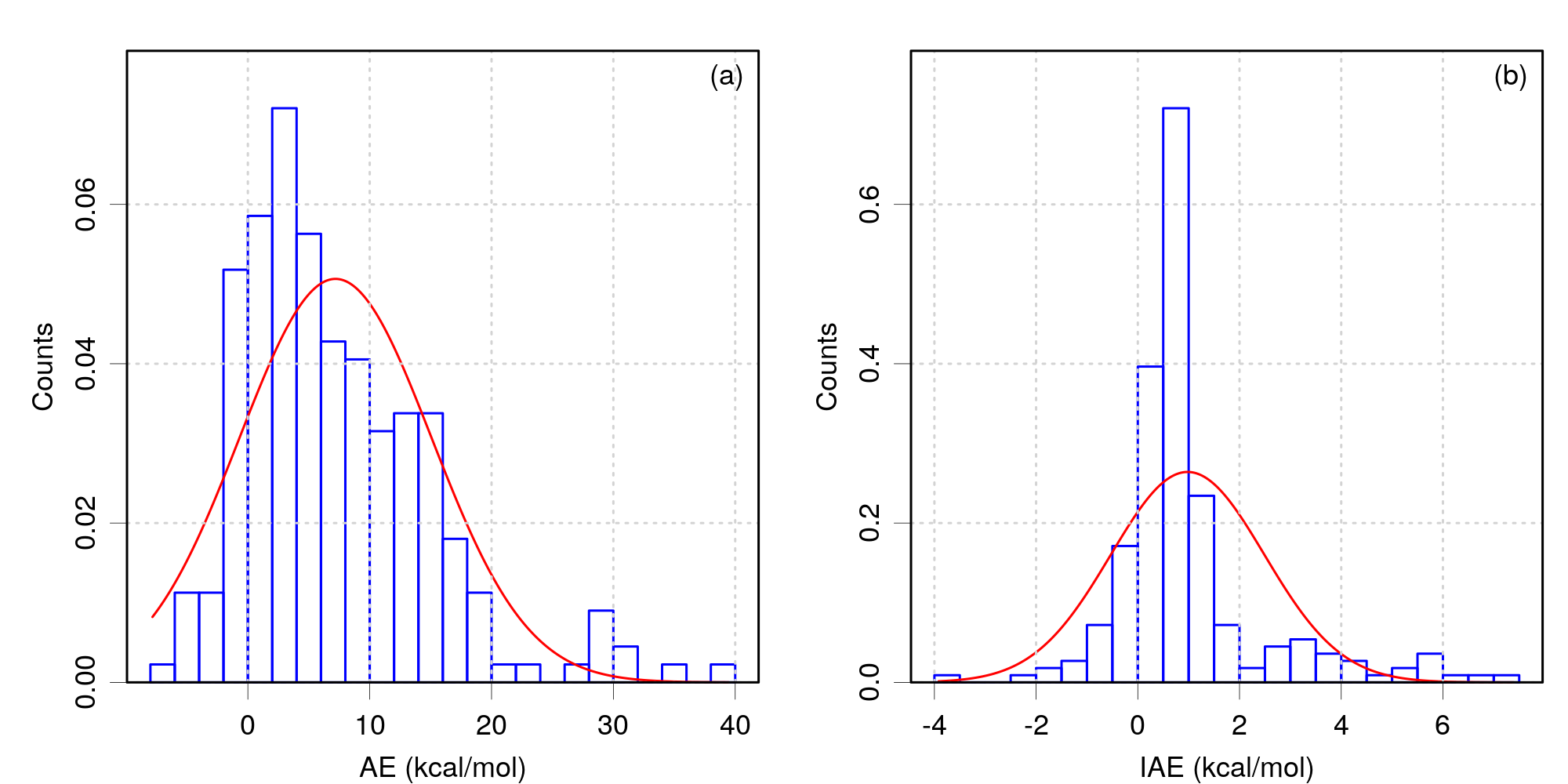}
\caption{\label{fig:fig1}\textbf{Histograms of the B3LYP errors for AE in
the G3/99 set}. A normal probability density function having the same
mean and standard deviation is superimposed. (a) atomization energies
($\mu=$7.2~kcal/mol, $\sigma=$7.9~kcal/mol); (b) intensive atomization
energies ($\mu=$1.0~kcal/mol, $\sigma=$1.5~kcal/mol). }
\end{figure}

Fig.~\ref{fig:fig1} shows histograms of the B3LYP errors. A normal
distribution having the same mean and standard deviation as the errors
set has been overlaid on the histogram. At a first glance, one notices
that the normal distribution does not faithfully describe the distribution
of errors. The latter has a more pronounced peak slightly right of
the origin, and presents some asymmetry: positive errors, even very
large ones, occur more often than negative ones. The deviation towards
positive errors explains why the normal distribution does not have
its center on the sharp peak, and also is broader than this peak.

Note that the non-normality observed on the histograms might also
be an effect of the limited size of the sample. Some numbers below
suggest, however, that this cannot be the only cause of discrepancy:
the sampling errors seem to be systematically lower than the discrepancies
one sees in Fig.~\ref{fig:fig1}. One is therefore in need of statistics
that convey useful information on non-normal distributions.

\subsubsection{Histograms do not tell the whole story}

\begin{figure}[!t]
\noindent \centering{}\includegraphics[width=0.45\textwidth]{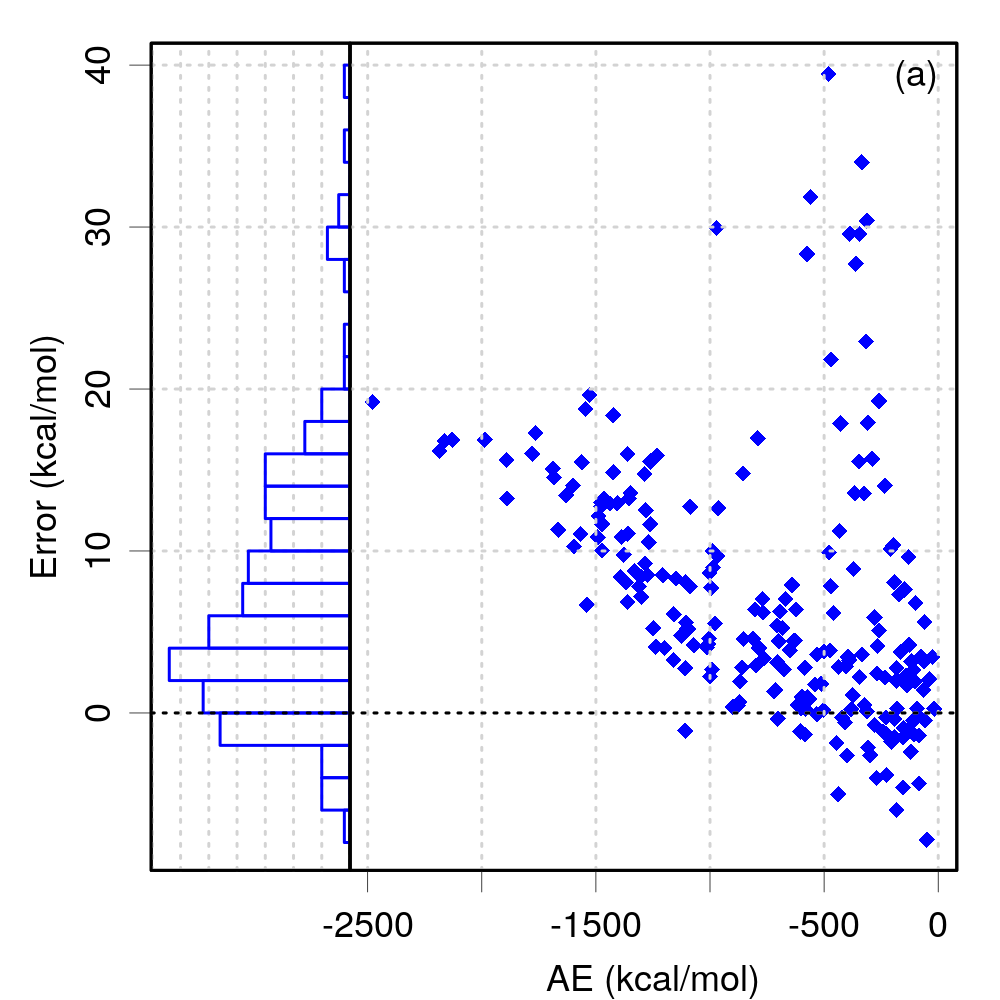}\includegraphics[width=0.45\textwidth]{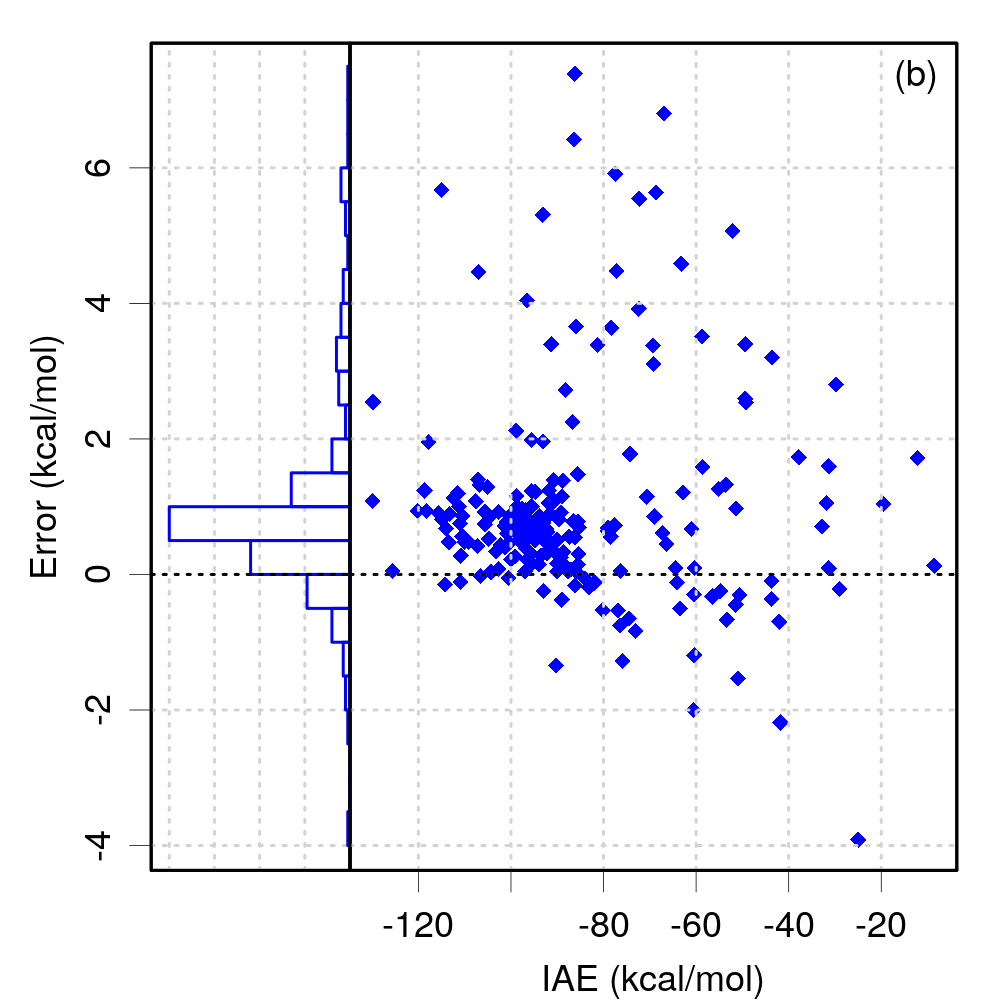}
\caption{\label{fig:fig1-1}\textbf{Distributions of the B3LYP errors for AE
and IAE in the G3/99 set}. Errors are plotted as a function of the
reference data. Histograms of the errors sets are plotted for comparison:
(a) atomization energies; (b) intensive atomization energies.}
\end{figure}

Histograms themselves are summaries that can hide important features
in the errors set. It is generally rewarding to analyze the errors
sample for underlying features, such as systems classes to be treated
separately \cite{Faver2011b}. Even histograms with a single maximum
(mode) can hide some heterogeneity in the sample. A very useful graphical
representation to reveal such features is to plot the errors as a
function of the calculated or reference property, as in Fig.~\ref{fig:fig1-1},
which displays side-by-side a scatterplot and the corresponding histogram.
The latter results from the projection and binning of the data cloud
on the ordinates axis: trends and heterogeneity in the data cloud
contribute to features in the histogram (asymmetry, multimodality,
...).

\begin{figure}[!t]
\noindent \centering{}\includegraphics[width=1\textwidth]{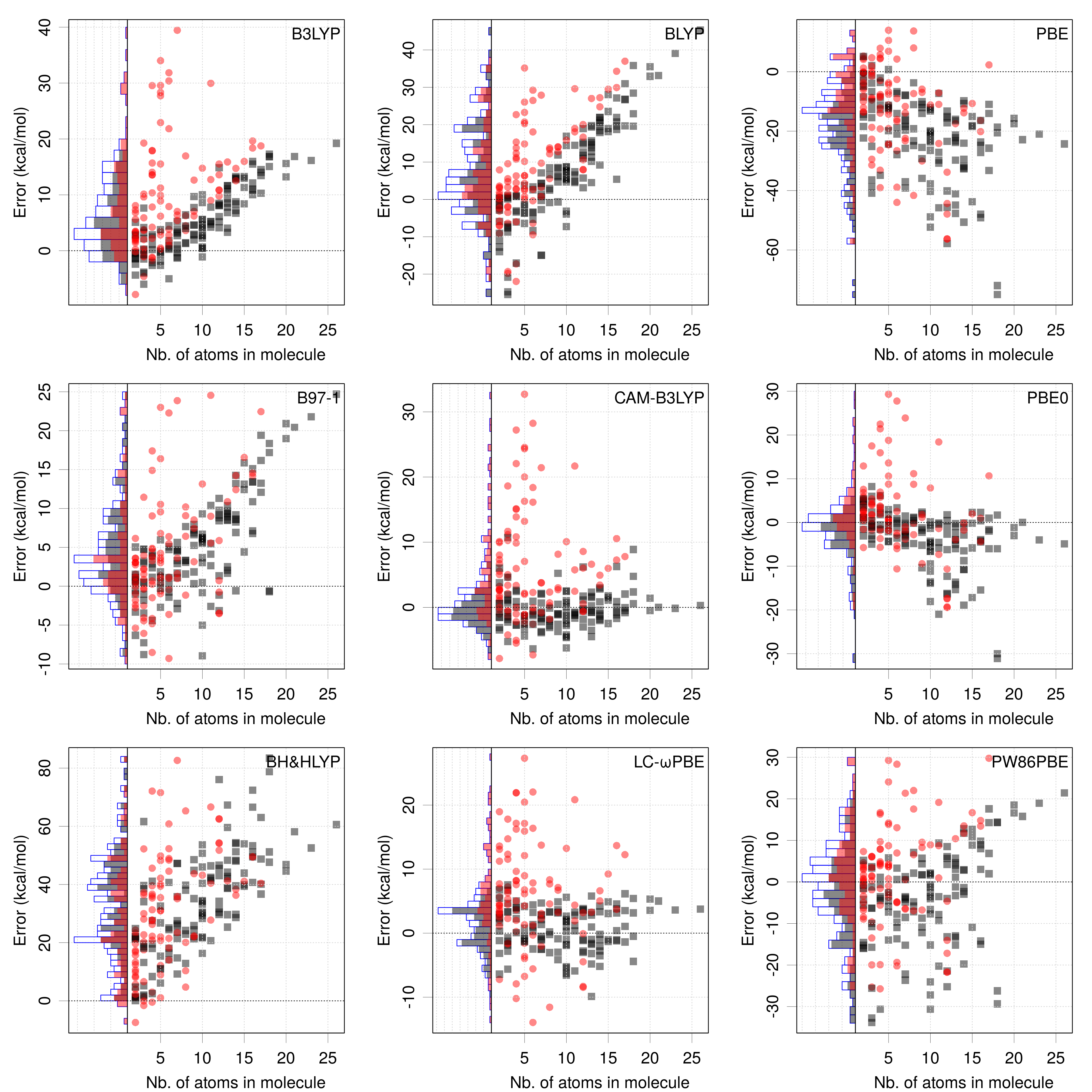}
\caption{\label{fig_Dist_AE}\textbf{Distribution of the errors for AE as a
function of the number of atoms in the molecules of the G3/99 set
for nine DFAs}. The data points are coded for CHON-type molecules
(gray squares and histogram) and the other ones (red circles and histogram).
The histograms corresponding to the partial and whole datasets are
displayed in the left panel of each graph. The histogram for the whole
dataset is traced with blue lines.}
\end{figure}

In the B3LYP case, one sees immediately that there are two problems:
(i) two branches in the dataset, with different trends, and (ii) a
strong (linear) dependence of the main set of errors with the atomization
energy. The upper, almost vertical, branch can be exclusively assigned
to molecules containing atoms out of the CHON set. The main, lower,
branch contains mostly CHON-type molecules, but also some non-CHON
systems. The linear trend in the main branch is linked to the extensivity
of atomization energies. This can be checked by plotting the errors
as a function of the number of atoms in the molecule (Fig.~\ref{fig_Dist_AE},
top left). The monotonous increase of the main branch with the number
of atoms is clear, whereas the effect is less marked for the non-CHON
branch. From this simple analysis, one sees that the prediction error
for an AE calculation with B3LYP will depend (1) on the nature of
the molecule, and (2) on its size.

Considering the error distribution for the other DFAs in Fig.~\ref{fig_Dist_AE},
different cases are observed: the linear increase of the AE errors
with the number of atoms is also observed for BLYP, BH\&HLYP and B97-1,
whereas CAM-B3LYP and LC-$\omega$PBE errors are mostly independent
of the molecule size, and an overall decrease is observed for PBE
and PBE0. The heterogeneity of non-CHON systems is mostly observed
for B3LYP, CAM-B3LYP, LC-$\omega$PBE, PBE0 and B97-1, whereas PBE,
PW86PBE and BH\&HLYP errors seem mostly uncorrelated with the chemical
composition.

To achieve the most accurate results for some DFAs, it would be desirable
to split the G3/99 set and perform statistics on the separate subsets.
However, for the sake of simplicity and fairness with regard to other
DFAs, we will continue here to work with the full G3/99 test set,
without questioning its homogeneity. 

\begin{figure}[!t]
\noindent \centering{}\includegraphics[width=1\textwidth]{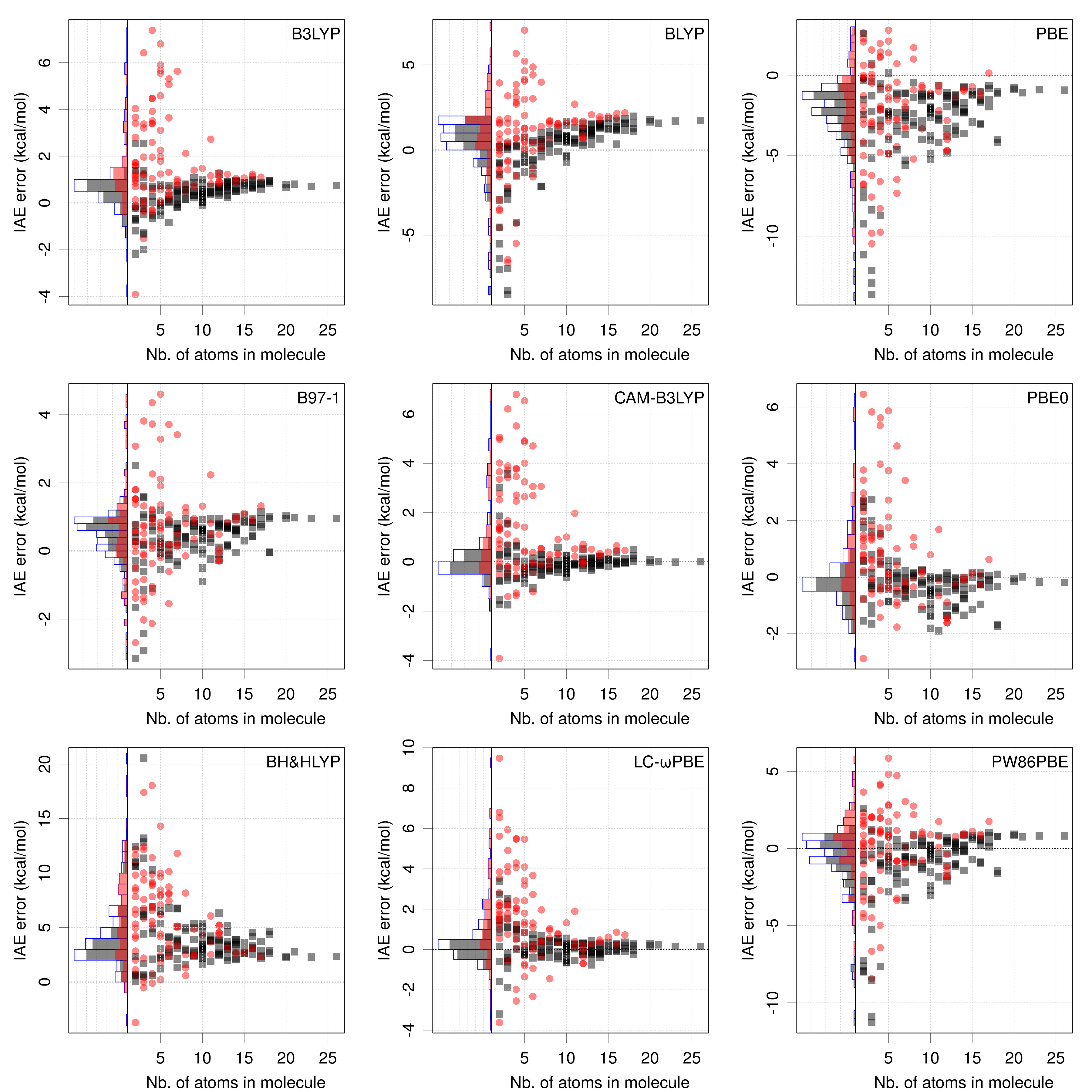}\caption{\label{fig_DIST_IAE}\textbf{Same as Fig.~\ref{fig_Dist_AE} for
IAE}.}
\end{figure}

The use of IAE solves in a large part the size-dependence problem
of AE (Fig.~\ref{fig_DIST_IAE}), but one is left with the composition
heterogeneity problem for some DFAs. Note that even for IAE, most
error distributions are neither normal, nor zero-centered (\emph{e.g.},
B3LYP, PBE, BLYP, BH\&HLYP, B97-1).

\subsubsection{Searching for outliers}

Points lying far in the wings of the histograms of error sets might
suggest inconsistent data in the reference set. Considering the linear
trends in the AE errors for several DFAs, extreme points might rather
be due to the molecule size than to a data problem. It is therefore
best to use IAE to identify outliers \cite{Perdew2016}. Outliers
have been tagged here as systems having IAE errors outside of the
95\,\% signed error range for a given DFA. The most common outliers
in the present DFA set are NO$_{2}$ (7/9 DFAs), SO$_{2}$, SiF$_{4}$,
N$_{2}$O, SO$_{3}$, O$_{2}$ (6/9 DFAs) and BeH (5/9 DFAs). Some
of these outliers have already been identified by Perdew \emph{et
al. }\cite{Perdew2016}, who discuss them with regard to the DFAs
properties.

\begin{figure}[!t]
\noindent \begin{centering}
\includegraphics[bb=0bp 0bp 1500bp 1200bp,clip,width=0.9\textwidth]{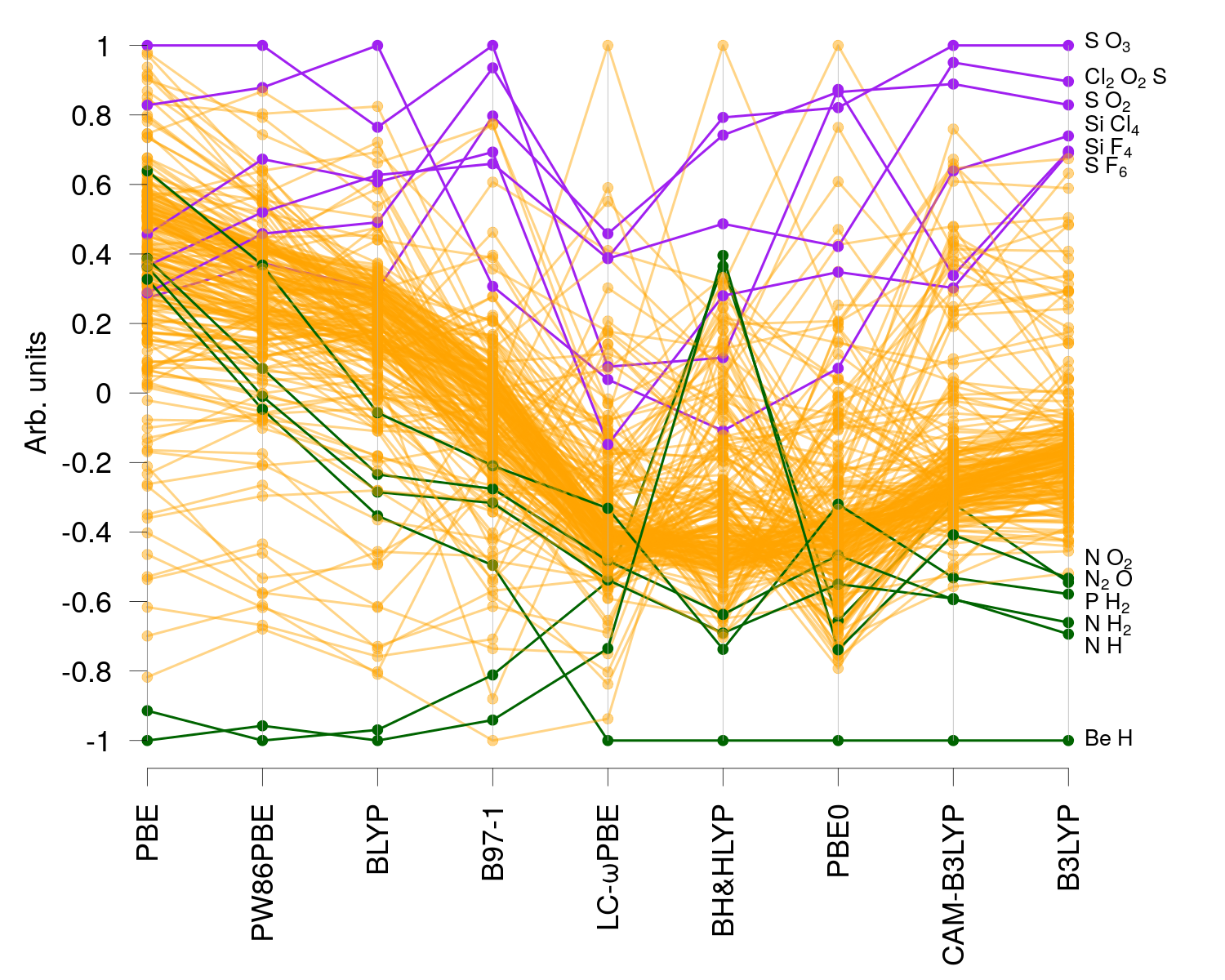} 
\par\end{centering}
\noindent \centering{}\caption{\label{fig_parout}\textbf{Parallel plot of the IAE errors in the
G3/99 dataset}. All error sets have been linearly rescaled to a common
$\left[-1,1\right]$ range. Lines join identical molecules in each
set. The DFAs have been ordered to display similarity in outliers
patterns. The labels on the right show selected outliers for B3LYP.}
\end{figure}

An important observation is that no outlier is common to all DFAs
(Fig.~\ref{fig_parout}), which indicates that the observed extreme
values are essentially due to limitations of the models, not to abnormal
reference data. There is therefore no solid reason to prune the dataset
in order improve the normality of the error distributions. As stated
above, one has definitely to deal with non-normal distributions and
adopt informative statistics enabling final users to make their choice
of DFA.

\subsubsection{Summary}

A useful tool to reveal features in the error sets is to plot the
errors as a function of the calculated or reference values, or any
other relevant property. Histograms contain more information than
summary statistics, but they do not tell the whole story!

From the exploration of the G3/99 dataset for AE and IAE, one might
underline that the error distributions are complex and structured
by several properties, such as the chemical composition (CHON \emph{vs}.
non-CHON) and the size of the molecule. Moreover, in these error sets,
the non-normality of the distributions is the rule rather than the
exception, which implies that the usual summary statistics are not
sufficient to enable reliable error predictions. This justifies the
need to turn to statistical tools not currently used in the CC methods
benchmarking literature, such as the cumulative probabilities and
percentiles presented in Section~\ref{sec:Probabilistic-statistics-of}.

Considering the size-dependence of AE errors for most DFAs in our
set (Fig.~\ref{fig_Dist_AE}), it is worthless to design simple and
reliable probabilistic indicators for this property. For instance,
B3LYP calculation for CHON molecules with more than 40 atoms will
present errors exceeding those present in the G3/99 set. In consequence,
only IAE error sets will be considered in the following.

\subsection{Probabilistic estimators for unsigned errors\label{sec:Results-and-discussion}}

In this section, we analyze the probabilistic estimators for unsigned
errors. Working on unsigned errors implies to accept a loss of information
to concentrate on the amplitude of errors. Note that probabilistic
estimators could as well be designed for signed errors, for instance
a pair of 2.5\% and 97.5\% quantiles delimiting a 95\,\% probability
interval, but they would lead to more complex ranking procedures.

\subsubsection{Statistics of unsigned errors and their uncertainty\label{sec:Statistics-of-unsigned} }

\begin{figure}[!t]
\noindent \centering{}\includegraphics[width=1\textwidth]{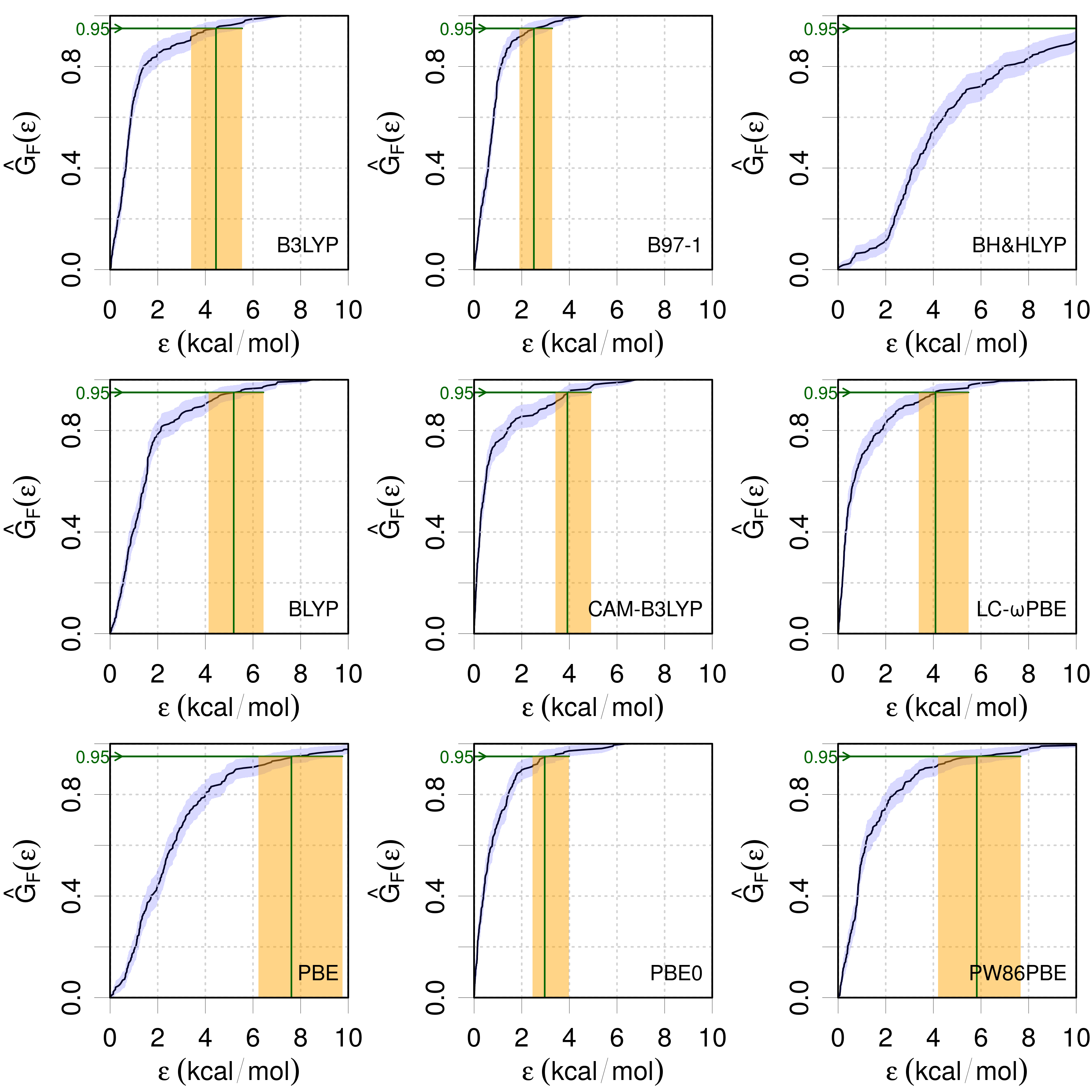}
\caption{\label{fig_ECDF_all_IAE}\textbf{Empirical cumulative distribution
function for unsigned errors on IAE, based on calculations for the
G3/99 set}. The shaded area delimits the 95\% uncertainty band on
the ECDF. The $Q_{95}$ percentile is indicated by a vertical green
line, and the orange area delimits its 95\,\% CI.}
\end{figure}

Statistics of unsigned IAE errors and their uncertainty have been
computed for all DFAs listed above: MUE, cumulative probability for
several thresholds, and a set of percentiles and limits of 95\,\%
CI for the higher percentiles. The uncertainties are reported in the
parenthetical notation, where ``the number in parentheses is the
numerical value of the standard uncertainty referred to the corresponding
last digits of the quoted result'' \cite{GUM}. The percentiles uncertainty
and CI limits have been calculated by bootstrapping, with 1000 repetitions
(Appendix~\ref{sec:Bootstraping}). The results are presented in
Table~\ref{tab:All-Stats}. The corresponding ECDFs are shown in
Fig.~\ref{fig_ECDF_all_IAE}, with the $Q_{95}$ percentile and its
95\,\% CI. Besides, these curves enable to estimate $C(\epsilon)$
and $Q_{n}$ at any level.
\begin{table}[!t]
\caption{\label{tab:All-Stats}\textbf{Statistics for the unsigned errors of
IAE for the G3/99 dataset}. The lower and upper limits of 95\,\%
confidence intervals on the $Q_{90}$and $Q_{95}$ percentiles are
notes as floor ($\left\lfloor x\right\rfloor $) and ceiling ($\left\lceil x\right\rceil $),
respectively. The optimal value in each column is noted by bold characters.}

\noindent \centering{}\medskip{}
\begin{tabular}{cr@{\extracolsep{0pt}.}lr@{\extracolsep{0pt}.}lr@{\extracolsep{0pt}.}lccr@{\extracolsep{0pt}.}lr@{\extracolsep{0pt}.}l}
\cline{1-11} 
{\footnotesize{}DFA} & \multicolumn{2}{c}{{\footnotesize{}$MUE$}} & \multicolumn{2}{c}{{\footnotesize{}$C(MUE)$}} & \multicolumn{2}{c}{{\footnotesize{}$C(0.25)$ }} & {\footnotesize{}$C(0.5)$ } & {\footnotesize{}$C(1.0)$} & \multicolumn{2}{c}{{\footnotesize{}$C(2.0)$}} & \multicolumn{2}{c}{}\tabularnewline
\cline{1-11} 
{\footnotesize{}B3LYP } & {\footnotesize{}1}&{\footnotesize{}2(1)} & {\footnotesize{}0}&{\footnotesize{}73(3) } & {\footnotesize{}0}&{\footnotesize{}15(2) } & {\footnotesize{}0.29(3) } & {\footnotesize{}0.68(3) } & {\footnotesize{}0}&{\footnotesize{}86(2) } & \multicolumn{2}{c}{}\tabularnewline
{\footnotesize{}B97-1 } & \textbf{\footnotesize{}0}&\textbf{\footnotesize{}85(5)} & {\footnotesize{}0}&{\footnotesize{}64(3) } & {\footnotesize{}0}&{\footnotesize{}20(3) } & {\footnotesize{}0.35(3) } & \textbf{\footnotesize{}0.75(3)}{\footnotesize{} } & \textbf{\footnotesize{}0}&\textbf{\footnotesize{}92(2) } & \multicolumn{2}{c}{}\tabularnewline
{\footnotesize{}BH\&HLYP } & {\footnotesize{}4}&{\footnotesize{}8(2)} & {\footnotesize{}0}&{\footnotesize{}64(3) } & {\footnotesize{}0}&{\footnotesize{}018(9) } & {\footnotesize{}0.02(1) } & {\footnotesize{}0.07(2) } & {\footnotesize{}0}&{\footnotesize{}12(2) } & \multicolumn{2}{c}{}\tabularnewline
{\footnotesize{}BLYP } & {\footnotesize{}1}&{\footnotesize{}6(1)} & {\footnotesize{}0}&{\footnotesize{}70(3) } & {\footnotesize{}0}&{\footnotesize{}08(2) } & {\footnotesize{}0.19(3) } & {\footnotesize{}0.41(3) } & {\footnotesize{}0}&{\footnotesize{}79(3) } & \multicolumn{2}{c}{}\tabularnewline
{\footnotesize{}CAM-B3LYP } & {\footnotesize{}0}&{\footnotesize{}90(9)} & \textbf{\footnotesize{}0}&\textbf{\footnotesize{}76(3}{\footnotesize{}) } & \textbf{\footnotesize{}0}&\textbf{\footnotesize{}41(3)}{\footnotesize{} } & \textbf{\footnotesize{}0.62(3)}{\footnotesize{} } & \textbf{\footnotesize{}0.76(3)}{\footnotesize{} } & {\footnotesize{}0}&{\footnotesize{}86(2) } & \multicolumn{2}{c}{}\tabularnewline
{\footnotesize{}LC-$\omega$PBE } & {\footnotesize{}1}&{\footnotesize{}1(1)} & {\footnotesize{}0}&{\footnotesize{}69(3) } & {\footnotesize{}0}&{\footnotesize{}30(3) } & {\footnotesize{}0.51(3) } & {\footnotesize{}0.69(3) } & {\footnotesize{}0}&{\footnotesize{}82(3) } & \multicolumn{2}{c}{}\tabularnewline
{\footnotesize{}PBE } & {\footnotesize{}2}&{\footnotesize{}8(2)} & {\footnotesize{}0}&{\footnotesize{}63(3) } & {\footnotesize{}0}&{\footnotesize{}04(1) } & {\footnotesize{}0.06(2) } & {\footnotesize{}0.18(3) } & {\footnotesize{}0}&{\footnotesize{}45(3) } & \multicolumn{2}{c}{}\tabularnewline
{\footnotesize{}PBE0 } & {\footnotesize{}0}&{\footnotesize{}92(8)} & {\footnotesize{}0}&{\footnotesize{}66(3) } & {\footnotesize{}0}&{\footnotesize{}30(3) } & {\footnotesize{}0.49(3) } & {\footnotesize{}0.68(3) } & {\footnotesize{}0}&{\footnotesize{}90(2) } & \multicolumn{2}{c}{}\tabularnewline
{\footnotesize{}PW86PBE } & {\footnotesize{}1}&{\footnotesize{}6(1)} & {\footnotesize{}0}&{\footnotesize{}69(3) } & {\footnotesize{}0}&{\footnotesize{}13(2) } & {\footnotesize{}0.24(3) } & {\footnotesize{}0.55(3) } & {\footnotesize{}0}&{\footnotesize{}75(3) } & \multicolumn{2}{c}{}\tabularnewline
\cline{1-11} 
\multicolumn{13}{c}{}\tabularnewline
\hline 
{\footnotesize{}DFA} & \multicolumn{2}{c}{{\footnotesize{}$Q_{50}$}} & \multicolumn{2}{c}{{\footnotesize{}$Q_{75}$}} & \multicolumn{2}{c}{{\footnotesize{}$Q_{90}$}} & \multicolumn{2}{c}{{\footnotesize{}$\left\lfloor Q_{90}\right\rfloor ,\left\lceil Q_{90}\right\rceil $}} & \multicolumn{2}{c}{{\footnotesize{}$Q_{95}$}} & \multicolumn{2}{c}{{\footnotesize{}$\left\lfloor Q_{95}\right\rfloor ,\left\lceil Q_{95}\right\rceil $}}\tabularnewline
\hline 
{\footnotesize{}B3LYP } & {\footnotesize{}0}&{\footnotesize{}80(4) } & {\footnotesize{}1}&{\footnotesize{}2(1) } & {\footnotesize{}3}&{\footnotesize{}2(5) } & \multicolumn{2}{c}{{\footnotesize{}2.0, 3.9 }} & {\footnotesize{}4}&{\footnotesize{}4(6) } & \multicolumn{2}{c}{{\footnotesize{}3.4, 5.5}}\tabularnewline
{\footnotesize{}B97-1 } & {\footnotesize{}0}&{\footnotesize{}70(4) } & \textbf{\footnotesize{}1}&\textbf{\footnotesize{}0(1)}{\footnotesize{} } & \textbf{\footnotesize{}1}&\textbf{\footnotesize{}6(2)}{\footnotesize{} } & \multicolumn{2}{c}{\textbf{\footnotesize{}1.4}{\footnotesize{}, }\textbf{\footnotesize{}2.1}{\footnotesize{} }} & \textbf{\footnotesize{}2}&\textbf{\footnotesize{}5(4)}{\footnotesize{} } & \multicolumn{2}{c}{\textbf{\footnotesize{}1.8}{\footnotesize{}, }\textbf{\footnotesize{}3.3}}\tabularnewline
{\footnotesize{}BH\&HLYP } & {\footnotesize{}3}&{\footnotesize{}8(2) } & {\footnotesize{}6}&{\footnotesize{}3(5) } & {\footnotesize{}10}&{\footnotesize{}0(7) } & \multicolumn{2}{c}{{\footnotesize{}8.4, 11.0 }} & {\footnotesize{}11}&{\footnotesize{}6(6) } & \multicolumn{2}{c}{{\footnotesize{}10.3, 12.4}}\tabularnewline
{\footnotesize{}BLYP } & {\footnotesize{}1}&{\footnotesize{}3(1) } & {\footnotesize{}1}&{\footnotesize{}8(1) } & {\footnotesize{}3}&{\footnotesize{}9(5) } & \multicolumn{2}{c}{{\footnotesize{}2.9, 4.6 }} & {\footnotesize{}5}&{\footnotesize{}2(6) } & \multicolumn{2}{c}{{\footnotesize{}4.3, 6.4}}\tabularnewline
{\footnotesize{}CAM-B3LYP } & \textbf{\footnotesize{}0}&\textbf{\footnotesize{}30(5)}{\footnotesize{} } & \textbf{\footnotesize{}0}&\textbf{\footnotesize{}9(2)}{\footnotesize{} } & {\footnotesize{}3}&{\footnotesize{}1(4) } & \multicolumn{2}{c}{{\footnotesize{}1.9, 3.8 }} & {\footnotesize{}3}&{\footnotesize{}9(4) } & \multicolumn{2}{c}{{\footnotesize{}3.4, 4.9}}\tabularnewline
{\footnotesize{}LC-$\omega$PBE } & {\footnotesize{}0}&{\footnotesize{}5(1) } & {\footnotesize{}1}&{\footnotesize{}4(2) } & {\footnotesize{}2}&{\footnotesize{}8(4) } & \multicolumn{2}{c}{{\footnotesize{}2.2, 3.8 }} & {\footnotesize{}4}&{\footnotesize{}1(6) } & \multicolumn{2}{c}{{\footnotesize{}3.4, 5.5}}\tabularnewline
{\footnotesize{}PBE } & {\footnotesize{}2}&{\footnotesize{}2(1) } & {\footnotesize{}3}&{\footnotesize{}5(3) } & {\footnotesize{}5}&{\footnotesize{}3(7) } & \multicolumn{2}{c}{{\footnotesize{}4.8, 7.1 }} & {\footnotesize{}7}&{\footnotesize{}6(9) } & \multicolumn{2}{c}{{\footnotesize{}6.4, 9.8}}\tabularnewline
{\footnotesize{}PBE0 } & {\footnotesize{}0}&{\footnotesize{}5(1) } & {\footnotesize{}1}&{\footnotesize{}3(1) } & {\footnotesize{}2}&{\footnotesize{}0(3) } & \multicolumn{2}{c}{{\footnotesize{}1.7, 2.7 }} & \textbf{\footnotesize{}3}&\textbf{\footnotesize{}0(5)}{\footnotesize{} } & \multicolumn{2}{c}{{\footnotesize{}2.5, 4.0}}\tabularnewline
{\footnotesize{}PW86PBE } & {\footnotesize{}0}&{\footnotesize{}9(1) } & {\footnotesize{}2}&{\footnotesize{}0(2) } & {\footnotesize{}3}&{\footnotesize{}6(5) } & \multicolumn{2}{c}{{\footnotesize{}2.8, 4.8 }} & {\footnotesize{}5}&{\footnotesize{}8(1) } & \multicolumn{2}{c}{{\footnotesize{}4.2, 7.7}}\tabularnewline
\hline 
\end{tabular}
\end{table}

If one considers the cumulative probabilities, several points are
outstanding. $C(\eta)$ for small values of $\eta$ (below the ``chemical
accuracy'' of 1 kcal/mol) are small for all DFAs, with a maximum
of 0.76(3) for CAM-B3LYP at $\eta=1$\,kcal/mol. Imposing smaller
error limits means accepting less reliable predictions. If one increases
the acceptance threshold to $\eta=2$\,kcal/mol, one reaches reasonable
confidence levels of 0.92(3) for B97-1 and 0.90(3) for PBE0. To achieve
the widely used confidence limit of 95\,\%, one has to accept higher
IAE error levels, for instance 3.9\,kcal/mol for CAM-B3LYP (\emph{cf.}
$Q_{95}$ values in Table~\ref{tab:All-Stats}).

The fact that, in order to make a statement that is valid with high
probability one has to accept large errors, is not conveyed by the
MUE. The latter might induce us to think that a \emph{typical} IAE
error level for methods such as B97-1, CAM-B3LYP and PBE0 is around
1\,kcal/mol. In fact, the cumulative probabilities at the MUE ($C(MUE)$
in Table~\ref{tab:All-Stats}) range between 0.63(3) and 0.76(3).
Note that this is higher than the upper limit of 0.5753 estimated
for the FND (Section~\ref{subsec:Cumulative-probabilities}; Fig.~\ref{fig_Q95}\,(a)),
but still low in terms of prediction confidence. In consequence, the
risk for the user to get absolute errors exceeding the MUE is unpredictable
from the MUE alone and rather high (up to 40\,\%). This disqualifies
the MUE as a basis for probabilistic estimations.

Looking at $Q_{95}$ one can see that three methods having similar
MUEs (B97-1, CAM-B3LYP and PBE0) can have significantly different
values of this high probability percentile, ranging from 2.5(4) for
B97-1 to 3.9(4)\,kcal/mol for CAM-B3LYP. This raises the interest
of $Q_{95}$ as a ranking metric, as reported below.

\subsubsection{Estimation of percentiles from MUE and MSE\label{subsec:Estimation-of-quantiles}}

\begin{figure}[!t]
\noindent \begin{centering}
\includegraphics[bb=0bp 0bp 1000bp 1000bp,width=0.45\textwidth]{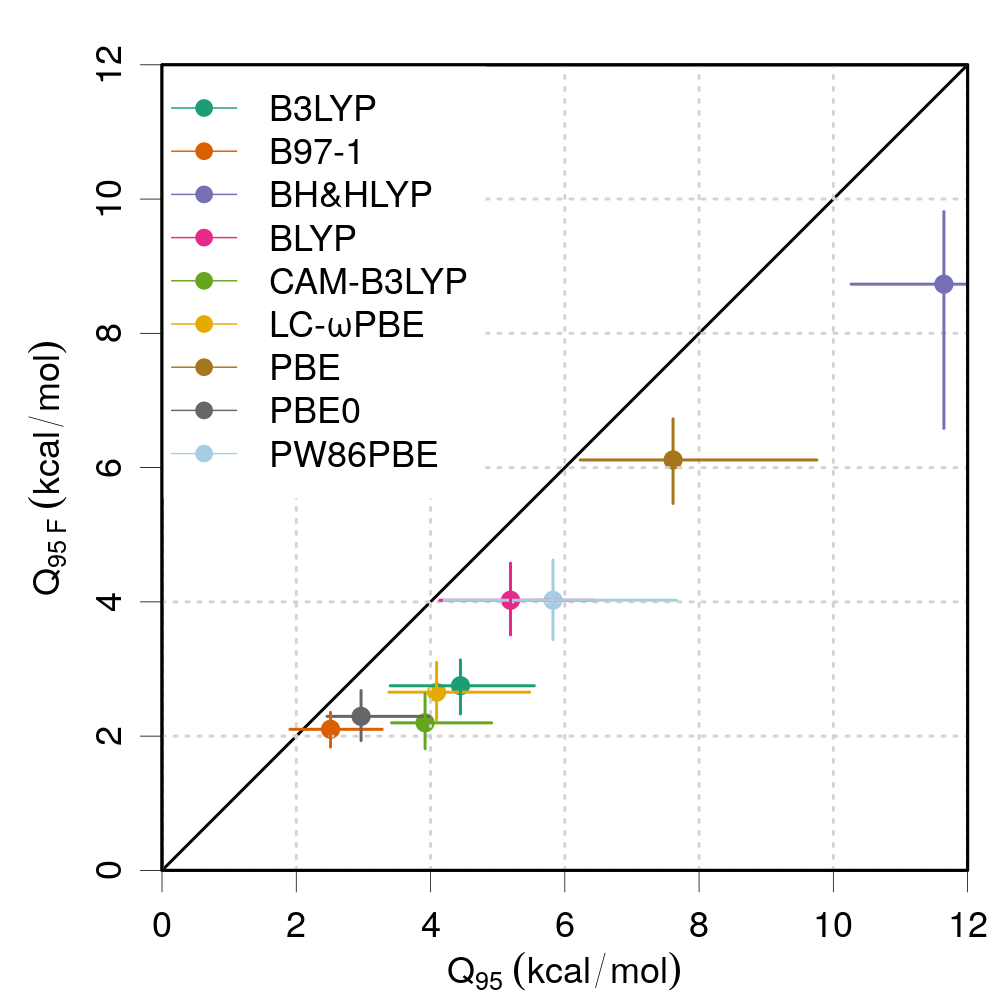}
\par\end{centering}
\noindent \centering{}\caption{\label{fig_Q95F}\textbf{Comparison of $Q_{95}$ and its approximation
$Q_{95F}$}. The error bars represent 95\,\% confidence intervals. }
\end{figure}

We have shown in Section~\ref{subsec:Quantiles} that, in the ideal
case of a normal error distribution, it is possible to estimate percentiles
of the corresponding folded distribution from MSE ($\hat{\mu}$) and
MUE ($\hat{\mu}_{F}$). This property is tested here on more realistic
error distributions. $Q_{95F}$ has been estimated from the MUE and
MSE, following the procedure described in Section \ref{subsec:Quantiles}.
A 95\,\% CI has been obtained by bootstrapping. Fig.~\ref{fig_Q95F}
compares $Q_{95F}$ and $Q_{95}$. Considering the position of the
points and the absence of intersection of the error bars with the
identity line, one can conclude that $Q_{95F}$ significantly underestimates
$Q_{95}$, except for B97-1, where the uncertainty on $Q_{95}$ is
large enough to leave a doubt. Due to the non-normality of the error
distributions, one cannot reliably estimate $Q_{n}$ from the generally
available MSE and MUE statistics.

\subsection{DFA ranking}

\begin{table}[t]
\noindent \centering{}\caption{\label{tab:Inversion-probabilities-1}\textbf{Inversion probabilities
in the MUE ranking}. These give the probability (as percentage) that
a row DFA achieves a lower MUE than a column DFAs with smaller mean
MUE value because of sampling uncertainty. The DFAs are ordered by
increasing mean MUE. Bold type indicates values higher than 20\,\%.}
\begin{tabular}{r@{\extracolsep{0pt}.}l|r@{\extracolsep{0pt}.}lr@{\extracolsep{0pt}.}lr@{\extracolsep{0pt}.}lr@{\extracolsep{0pt}.}lr@{\extracolsep{0pt}.}lr@{\extracolsep{0pt}.}lr@{\extracolsep{0pt}.}lr@{\extracolsep{0pt}.}l}
\multicolumn{2}{c|}{} & \multicolumn{2}{c}{B97-1 } & \multicolumn{2}{c}{CAM-B3LYP } & \multicolumn{2}{c}{PBE0 } & \multicolumn{2}{c}{LC-$\omega$PBE } & \multicolumn{2}{c}{B3LYP } & \multicolumn{2}{c}{BLYP } & \multicolumn{2}{c}{PW86PBE } & \multicolumn{2}{c}{PBE }\tabularnewline
\hline 
\multicolumn{2}{c|}{B97-1} & \multicolumn{2}{c}{} & \multicolumn{2}{c}{} & \multicolumn{2}{c}{} & \multicolumn{2}{c}{} & \multicolumn{2}{c}{} & \multicolumn{2}{c}{} & \multicolumn{2}{c}{} & \multicolumn{2}{c}{}\tabularnewline
\multicolumn{2}{c|}{CAM-B3LYP } & \multicolumn{2}{c}{\textbf{44 }} & \multicolumn{2}{c}{} & \multicolumn{2}{c}{} & \multicolumn{2}{c}{} & \multicolumn{2}{c}{} & \multicolumn{2}{c}{} & \multicolumn{2}{c}{} & \multicolumn{2}{c}{}\tabularnewline
\multicolumn{2}{c|}{PBE0 } & \multicolumn{2}{c}{\textbf{40 }} & \multicolumn{2}{c}{\textbf{47 }} & \multicolumn{2}{c}{} & \multicolumn{2}{c}{} & \multicolumn{2}{c}{} & \multicolumn{2}{c}{} & \multicolumn{2}{c}{} & \multicolumn{2}{c}{}\tabularnewline
\multicolumn{2}{c|}{LC-$\omega$PBE } & \multicolumn{2}{c}{\textbf{23 }} & \multicolumn{2}{c}{\textbf{28 }} & \multicolumn{2}{c}{\textbf{30 }} & \multicolumn{2}{c}{} & \multicolumn{2}{c}{} & \multicolumn{2}{c}{} & \multicolumn{2}{c}{} & \multicolumn{2}{c}{}\tabularnewline
\multicolumn{2}{c|}{B3LYP } & \multicolumn{2}{c}{14 } & \multicolumn{2}{c}{19 } & \multicolumn{2}{c}{\textbf{20 }} & \multicolumn{2}{c}{\textbf{39 }} & \multicolumn{2}{c}{} & \multicolumn{2}{c}{} & \multicolumn{2}{c}{} & \multicolumn{2}{c}{}\tabularnewline
\multicolumn{2}{c|}{BLYP } & \multicolumn{2}{c}{1 } & \multicolumn{2}{c}{2 } & \multicolumn{2}{c}{2 } & \multicolumn{2}{c}{6 } & \multicolumn{2}{c}{9 } & \multicolumn{2}{c}{} & \multicolumn{2}{c}{} & \multicolumn{2}{c}{}\tabularnewline
\multicolumn{2}{c|}{PW86PBE } & \multicolumn{2}{c}{2 } & \multicolumn{2}{c}{2 } & \multicolumn{2}{c}{3 } & \multicolumn{2}{c}{7 } & \multicolumn{2}{c}{11 } & \multicolumn{2}{c}{\textbf{49} } & \multicolumn{2}{c}{} & \multicolumn{2}{c}{}\tabularnewline
\multicolumn{2}{c|}{PBE } & \multicolumn{2}{c}{0 } & \multicolumn{2}{c}{0 } & \multicolumn{2}{c}{0 } & \multicolumn{2}{c}{0 } & \multicolumn{2}{c}{0 } & \multicolumn{2}{c}{0 } & \multicolumn{2}{c}{0 } & \multicolumn{2}{c}{}\tabularnewline
\multicolumn{2}{c|}{BH\&HLYP } & \multicolumn{2}{c}{0 } & \multicolumn{2}{c}{0 } & \multicolumn{2}{c}{0 } & \multicolumn{2}{c}{0 } & \multicolumn{2}{c}{0 } & \multicolumn{2}{c}{0 } & \multicolumn{2}{c}{0 } & \multicolumn{2}{c}{0 }\tabularnewline
\end{tabular}
\end{table}

\subsubsection{Impact of statistical uncertainty on MUE-based ranking\label{subsec:Pairwise-comparisons-based}}

When ranking DFAs by their MUE, the sampling uncertainty on the statistic
has ideally to be taken into account, which, to our knowledge, is
never reported in the literature. 

Considering the MUE for two DFAs, $MUE_{1}$ and $MUE_{2}$ with mean
values and standard errors $\mu_{F1}\pm u_{1}$ and $\mu_{F2}\pm u_{2}$
(Eq.~\ref{eq:uMUE}), the probability density function of $MUE_{1}-MUE_{2}$
is a normal PDF with mean $\mu=\mu_{F1}-\mu_{F2}$ and variance $\sigma^{2}=u_{1}^{2}+u_{2}^{2}$.
Therefore, one gets $P(MUE_{1}-MUE_{2}<0)$ as the cumulative probability
\begin{equation}
P(MUE_{1}<MUE_{2})\simeq\Phi(0,\mu=\mu_{F1}-\mu_{F2},\sigma^{2}=u_{1}^{2}+u_{2}^{2})\label{eq:inversP}
\end{equation}
where $\Phi(x;\mu,\sigma^{2})$ is the cumulative distribution function
for a normal distribution with mean $\mu$ and variance $\sigma^{2}$
(\emph{cf.} Section \ref{subsec:Statistical-uncertainty-of}). Using
Eq.~\ref{eq:inversP}, an \emph{ordering inversion probability} has
been evaluated for pairs of DFAs with $\mu_{F1}>\mu_{F2}$, and reported
in Table\,\ref{tab:Inversion-probabilities-1}. Note that this configuration
implies that the upper limit of the inversion probability is $0.5$. 

There is a neat segregation of the DFAs in two groups: (1) B97-1,
CAM-B3LYP, PBE0, LC-$\omega$PBE and B3LYP, among which the inversion
risk is medium to very high; and (2) BLYP, PW86PBE, PBE and BH\&HLYP,
which have vanishing chances to outperform any DFA of the first group.
In the second group, the MUE ranking of PW86PBE and BLYP is not statistically
significant.

\subsubsection{Ranking by percentiles}

As we have ruled out the use of MUE for probabilistic estimation,
could it also be replaced for DFA ranking? Ranking of approximations
could be done according to the values of $C(\epsilon)$ for a given
$\epsilon$: the higher $C(\epsilon)$ the better the method. Alternatively,
one can rank approximations by choosing a percentile $Q_{n}$: the
lower $Q_{n}$, the better the method. As one can more easily and
generally agree on a reference percentile than on an error level,
the former being independent on the type of analyzed property, we
test here how high-probability percentiles such as $Q_{95}$ can be
used for the ranking of DFAs, and how they compare to MUE-based ranking.
\begin{figure}[!t]
\noindent \centering{}\includegraphics[bb=0bp 0bp 2000bp 1200bp,width=0.9\textwidth]{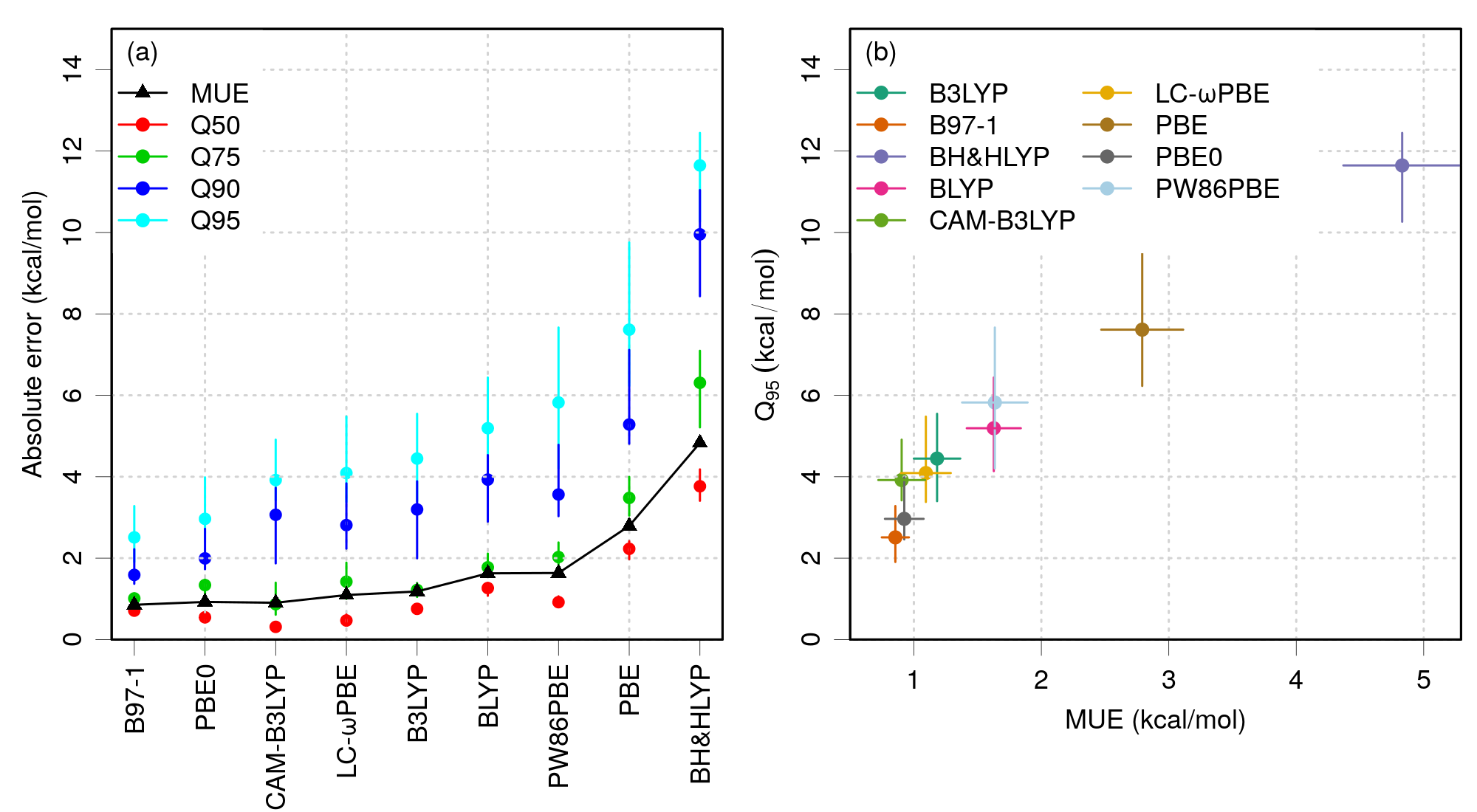}
\caption{\label{fig_Q95}\textbf{Comparison of statistics for ranking.} (a)
$MUE$ and $Q_{50}$, $Q_{75}$, $Q_{90}$ ,$Q_{95}$ percentiles
for the set of DFAs, sorted by increasing value of  $Q_{95}$; (b)
correlation between \emph{MUE} and $Q_{95}$. The error bars represent
95\,\% confidence intervals. }
\end{figure}

In order to facilitate the comparison between methods, the percentiles
in Table~\ref{tab:All-Stats} have been plotted together in Fig.~\ref{fig_Q95}(a),
along with the MUE, and sorted by increasing $Q_{95}$ values. One
sees that CAM-B3LYP is best at the 50\,\% level, but is challenged
by B97-1 at the 75\,\% level and then by PBE0 at higher probability
levels. As noted above, CAM-B3LYP, B97-1 and PBE0 have the same $MUE$
for this property ($\sim0.9(1)$). The high percentiles can thus provide
additional discriminating ranking criteria. Note that it is not surprising
that, as a rule, hybrid methods (with the notable exception of the
pioneering BH\&HLYP) come out better than pure functionals. 

Globally, if one compares the ranking by $MUE$ and $Q_{95}$ (Fig.~\ref{fig_Q95}(b)),
the correlation is strong, except for the CAM-B3LYP/PBE0 inversion,
which is not statistically significant, considering the high inversion
probability estimated from the MUE standard errors (Table~\ref{tab:Inversion-probabilities-1}).
The consideration of the error bars on the statistics shows that a
strict ranking by the mean value of the statistics is not pertinent
here. The definition of groups of methods would be more statistically
relevant. 

So, it appears that $Q_{95}$, beyond its added value for prediction
errors estimation, would also be a relevant substitute to $MUE$ for
the ranking of DFAs, or any computational chemistry methods. On the
present dataset, it does not profoundly scramble the usual ranking,
which is a reassuring point for its introduction in future benchmarks. 

\section{Conclusion}

Although testing computational chemistry methods on reliable data
sets is nowadays the preferred validation method, finding relevant
measures to validate and rank them it is still an open problem. One
of the difficulties is that the distributions of errors for uncorrected
method are often far from a normal distribution. They are typically
asymmetric, not zero-centered, and correlated, which precludes the
estimation of a prediction uncertainty, \emph{i.e.}, without applying
corrections of systematic errors. Even for normally distributed errors,
the \emph{unsigned} errors are not normally distributed, but follow
the so-called ``folded normal distribution'' (Fig.~\ref{fig_folded}\,(a)).
One should thus avoid thinking about a normal distribution when analyzing
unsigned errors. Their mean value ($MUE$) is neither close to the
mode of the unsigned error distribution, nor to its median.

An important aspect of the present study is the assessment of the
statistical uncertainty on the estimators due to the limited size
of reference data sets, and the illustration of their impact on the
conclusions that are drawn from them. For instance, the rank differences
between some methods are not significant in view of the ranking statistics
uncertainties. Although the error sets cannot generally be assumed
to be uncorrelated, we recommend that the standard errors of the statistics
should be systematically published. These standard errors are most
certainly underestimated, but they still can be useful to assess the
statistical reliability of rankings.

We have shown that, because of the non-normality of the error distributions,
the MUE cannot be used to communicate probabilistic statements. In
the examples and error samples studied here, the probability that
absolute errors exceed the MUE range from 0.2 to 0.5. In consequence,
we propose to use estimators based on the empirical cumulative distribution
function (ECDF) of the \emph{unsigned} errors: the cumulative probabilities
$C(\eta)$ and the percentiles $Q_{n}$. They can be used in two typical
scenarii: 
\begin{itemize}
\item the end-users choose first a value $\eta$ of the maximal admissible
absolute error for their application, and obtain from the reference
data set an estimate of the percentage of acceptable results for a
given method at this error level, $C(\eta)$; or
\item the users choose a percentage of acceptable results required for their
application ($n$\,\%) or a risk level $(100-n)$\,\%, and get the
maximal error they have to accept when using a given method, $Q_{n}$. 
\end{itemize}
In the latter case, one is typically interested in high percentages,
such as $n=90$\,\% or 95\,\%, the latter being preferred in order
to link with the recommended usage in thermochemistry to report an
enlarged uncertainty $u_{95\%}$ \cite{Ruscic2014}.\texttt{ }We have
seen that, due to the shape of error distributions, high-probability
percentiles, such as $Q_{95}$, cannot be reliably estimated from
the usual statistics (MSE, MUE, RMSE...). Besides, we have shown that,
for the end-user, they convey much more useful information than the
MUE, and also that they provide similar methods rankings as the latter.
We therefore recommend that $Q_{95}$ percentiles should be tabulated
in addition to the conventional statistics, along with their standard
errors. Systematic publication of the ECDF curves could also be a
very interesting addition.

There are a few \emph{caveats} on the use of probabilistic estimators.
They should not be used for error sets where there is a notable trend,
such as the molecule size dependence known for the atomization energies.
In this case, all calculated values for molecules larger than the
ones in the reference set are expected to have errors beyond the estimated
$Q_{95}$, breaking the probabilistic interpretation and usefulness
of the latter. The second \emph{caveat} concerns the size of the reference
dataset. The uncertainty in the high percentiles increases rapidly
as the set size decreases. It is probably not reasonable to trust
a $Q_{95}$ value for datasets with less than typically 100 points
(See Appendix \ref{sec:Bootstraping}). In any case, the confidence
limits on the percentiles should be estimated, for instance by bootstrapping
techniques.

The calculation of the $C$-type estimators depends on the users choice
of an application-dependent acceptable error level, and therefore
cannot be easily tabulated, or maybe for some typical error values
(chemical accuracy...). It is therefore desirable that reference databases
provide an easy access to error data and tools to extract and treat
them. This would make it more easy to the end-users to make a rational
and informed choice of method. On a more general basis, authors of
benchmarking/ranking studies should aim at reproducibility, and provide
their error datasets in \emph{machine-readable} format (\emph{e.g.},
in tabular form, as supplementary material \texttt{.csv} files) \cite{Walters2013,Rudshteyn2017}.
Data recovery from tables in \texttt{pdf} files often requires error-prone
human post-treatment, notably when the data tables are rotated or
contain empty cells, references as superscripts, or typographical
minus signs for negative numbers.

Although we do not intend to make recommendations for, or against
a given DFA, the present results confirm the widespread opinion that
hybrids are, in many cases, superior to pure functionals. We have
also seen that the performances of the studied density functionals
are not very high\footnote{Note that the performances of the studied DFAs for atomization energies
could have been significantly improved in two ways: (1) by splitting
the G3/99 datasets into CHON and non-CHON subsets; and (2) by correction
of the bias and/or trends observed in the error samples \cite{Pernot2015,Ramakrishnan2015}.
At the present level of ``raw errors'', the percentiles of the unsigned
error distributions include prediction bias ($MSE\ne0$). They provide
estimates of the expected error amplitude which are therefore pessimistic,
in the sense that a simple shift of the results by the MSE would notably
improve the situation for many DFAs. }. However, some of them are widely used and appreciated. Could it
be that the need for high accuracy is often exaggerated? Let us consider
that, even if the chemical accuracy is far to be reached for AE, this
does not prevent more accurate results to be generated for reaction
enthalpies, thanks to error cancellations \cite{Margraf2017}. Moreover,
it has been repeatedly shown that reliable conclusions on catalytic
and surface reactions can be drawn from moderately accurate DFT calculations,
provided prediction uncertainties and their correlations are carefully
estimated and accounted for \cite{Medford2014,Sutton2016,Ulissi2017}. 
\begin{acknowledgments}
The authors would like to thank Erin Johnson for providing the dataset
of AE calculations for this study.
\end{acknowledgments}

\section*{Supplementary Material}

See supplementary material for access to datasets and \texttt{R} code
for data analysis and generation of figures and tables of the article
\cite{ECDFT}.

\bibliographystyle{aipnum4-1}

\appendix
\part*{Appendices}

\section{Estimation of percentiles CI by bootstrapping\label{sec:Bootstraping}}

\begin{figure}[!t]
\noindent \centering{}\includegraphics[width=0.9\textwidth]{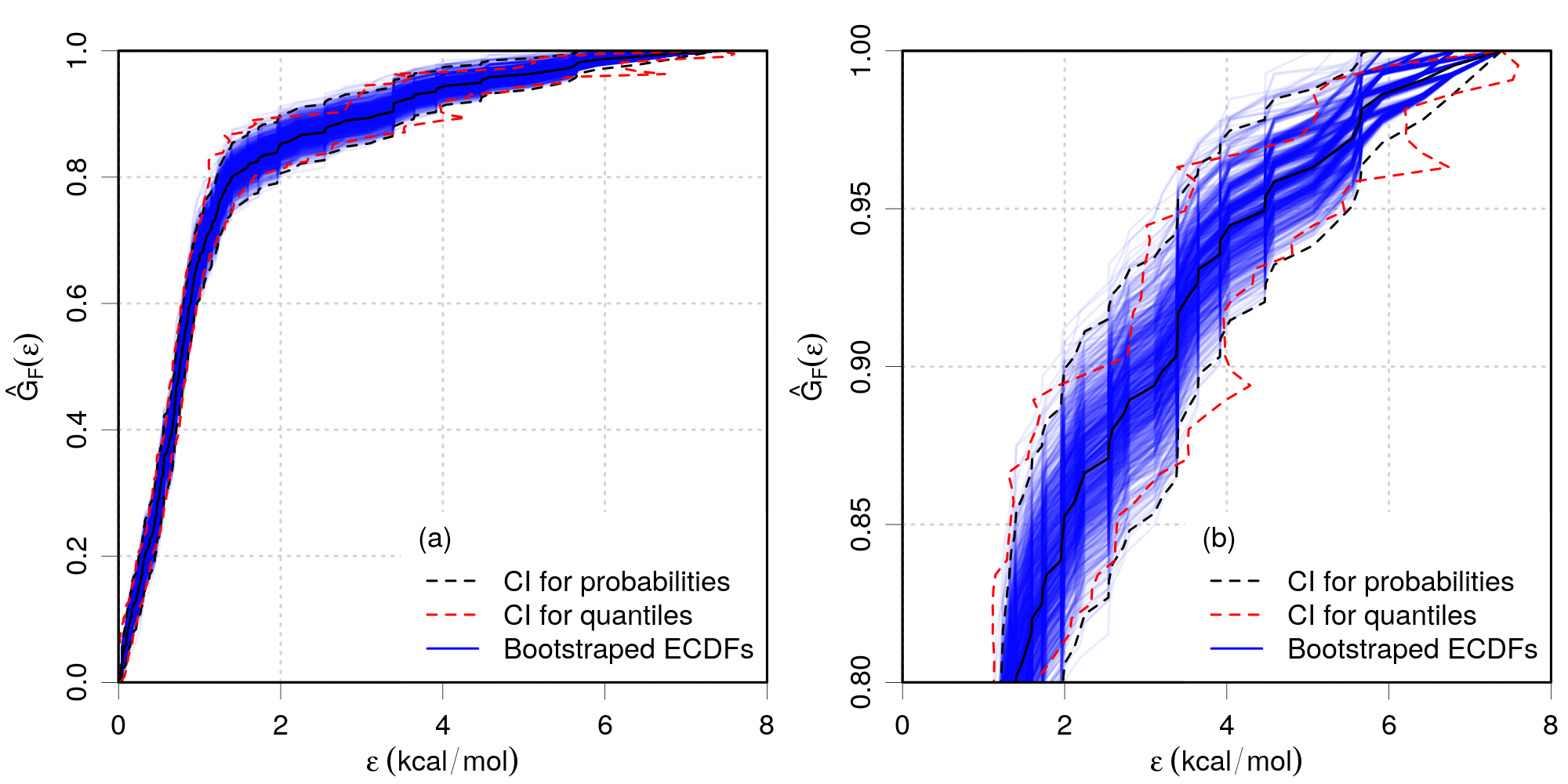}\caption{\label{fig:Verification-of-formulae}\textbf{Verification of formulae
for vertical and horizontal uncertainties on the ECDF for IAE errors
by B3LYP}. (a) full probability range; (b) closeup on the high probability
range.}
\end{figure}

The uncertainty of percentiles $Q_{n}$ has been estimated by Kendall's
formula (Eq.~\ref{eq:uQ}) and by bootstrapping of the B3LYP errors
for IAE. To compute Kendall's formula, an estimation of the probability
density function $\pi_{F}$ has been generated by a kernel density
method (\texttt{density()} function of the \texttt{R} language \cite{RTeam2015}).
95\,\% confidence intervals have been approximated by a normal enlargement
factor ($\pm1.96\times u_{Q_{P}}$) and plotted in Fig.~\ref{fig:Verification-of-formulae}
(red dashed curves). 

A sample of 1000 bootstrapped errors sets has been generated by random
sampling the original errors set with replacement. From this sample
of error sets, ECDFs have been plotted as reference in Fig.~\ref{fig:Verification-of-formulae}
(blue curves), and 95\,\% confidence intervals have been estimated
for all percentiles. These CIs have been plotted in Fig.~\ref{fig:Verification-of-formulae}
(black dashed curves). They are indistinguishable of the confidence
limits on the cumulative probabilities $C(\epsilon)$ obtained by
Wald's formula (Eq.~\ref{eq:uPr}).

By contrast, the CI on $Q_{95}$ (red-dashed) starts to deviated notably
from the reference CI above $p\simeq0.8$ (Fig.~\ref{fig:Verification-of-formulae}(b)).
Even with a fairly large error sample ($N=222$), the estimation of
the tails of $\pi_{F}$ cannot be relied upon for use in Kendall's
formula. The quantiles uncertainty and confidence limits are better
estimated by bootstrap, in which case they are consistent with the
cumulative probabilities uncertainty estimated by Wald's formula.
These values for $Q_{50}$, $Q_{90}$ and $Q_{95}$ are reported in
Table~\ref{tab:All-Stats}, for all DFAs.

\paragraph{Sample size effect.}

\begin{figure}[!t]
\noindent \begin{centering}
\includegraphics[width=0.9\textwidth]{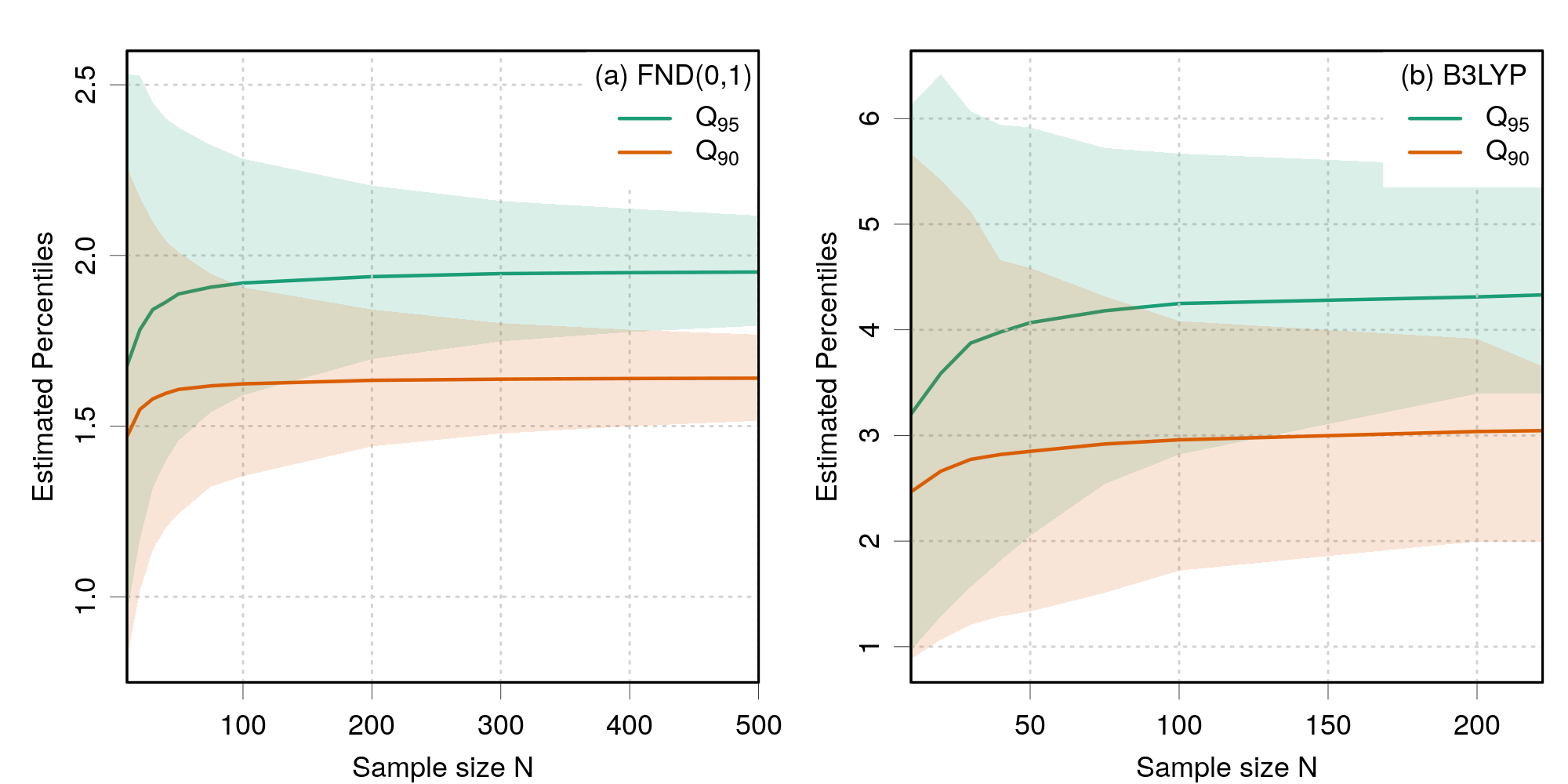}
\par\end{centering}
\caption{\label{fig:Convergence-with-the}\textbf{Convergence with the sample
size $N$, of the estimated percentiles $Q_{90}$ and} $Q_{95}$.
(a) folded normal distribution $\pi_{FN}(\epsilon;\mu=0,\sigma=1)$,
noted $FND(0,1)$; (b) subsets of the B3LYP error set. Full lines
represent the mean value of the percentiles and shaded areas delimit
95\,\% confidence intervals.}
\end{figure}

The results above raise the question of the impact of the sample size
on the CI limits of high percentiles. In order to appreciate this
effect, we performed a Monte Carlo study by generating 10000 random
samples of the folded normal distribution $\pi_{FN}(\epsilon;\mu=0,\sigma=1)$,
for sizes between $N=10$ to $500$. For each value of $N$, the mean
and 95\,\% confidence limits of $Q_{90}$ and $Q_{95}$ have been
estimated from the sample of 10000 values. The corresponding curves
are shown in Fig.~\ref{fig:Convergence-with-the}(a).

Below $N=100$, there is a strong overlap of the distributions, in
the sense that the mean value of one percentile lies within the 95\,\%
CI of the other. Above this value, there is a better discrimination,
but one has to wait until $N\simeq400$ to get non-overlapping 95\,\%
CI intervals. 

A similar plot has been done by bootstrapping subsets of the B3LYP
data to evaluate the effect of the non-normality of the error distribution
on this analysis. One sees in Fig.~\ref{fig:Convergence-with-the}(b)
that the conclusions are similar: indiscernibility of $Q_{90}$ and
$Q_{95}$ below $N\simeq100$, with a small overlap of the 95\,\%
CIs around $N=200$.
\end{document}